\documentclass[aps,twocolumn,groupedaddress,showpacs]{revtex4-1}
\usepackage{graphicx,amsfonts,amssymb,amsmath,subfigure}
\begin{document}


\title{Comment on ``Observation of two-dimensional Anderson localisation of ultracold atoms''}



\author{Sophie S.~Shamailov}
\email[]{sophie.s.s@hotmail.com}
\affiliation{Department of Physics, University of Auckland, Private Bag 92019, Auckland 1142, New Zealand}



\date{\today}

\begin{abstract}
Here I provide additional experimental information and criticise the analysis, modelling, interpretation and claims presented in the recent paper \cite{BS}. I argue that the authors have no clear experimental evidence of Anderson localisation (as they claim) and their numerical simulations are very far indeed from reproducing the experiment, as stated in their article. Furthermore, some effects that are ascribed to real physical mechanisms are revealed to be simply artefacts.
\end{abstract}

\pacs{}

\maketitle
%
\section{Introduction}
After multiple discussions with both the experimental and theoretical groups that have authored \cite{BS}, I have formally requested permission to present the experimental data (fully acknowledging its rightful owners, of course), and believe I have important information to bring to light. The raw data that was included in the publication \cite{BS} is now publicly available through an online repository \cite{repo1}. All experimental data shown in this Comment is taken from \cite{repo1} and belongs to the authors of \cite{BS,repo1}. However, there is considerable additional \textit{un}published data (including, but not limited to, that shown in the thesis \cite{ThomasPhD}), which I find to be very relevant to interpreting the published results, and which is currently not publicly available or citable. Should this change in the future, several arguments put forward below could be considerably strengthened on the basis of this data. For future reference, please note that the fill factor in \cite{ThomasPhD} was accidentally quoted as four times less than the actual preset value (not accounting for scatterer overlap) used in the experiments.

The purpose of this Comment is to present all the relevant experimental evidence (restricted to the publicly available data), elucidate the shortcomings of the simulations and analysis performed in \cite{BS}, and argue that the claims made therein are unjustified. However, I encourage readers to think and judge for themselves, not taking either point of view or argument for granted. The rest of the Comment is written for an audience that is already well familiar with the content of \cite{BS}.
\section{Experimental information}
\subsection{Fringe, acceleration and initial condition}
\label{FAIC}
\begin{figure}[htbp]
\includegraphics[width=3.45in]{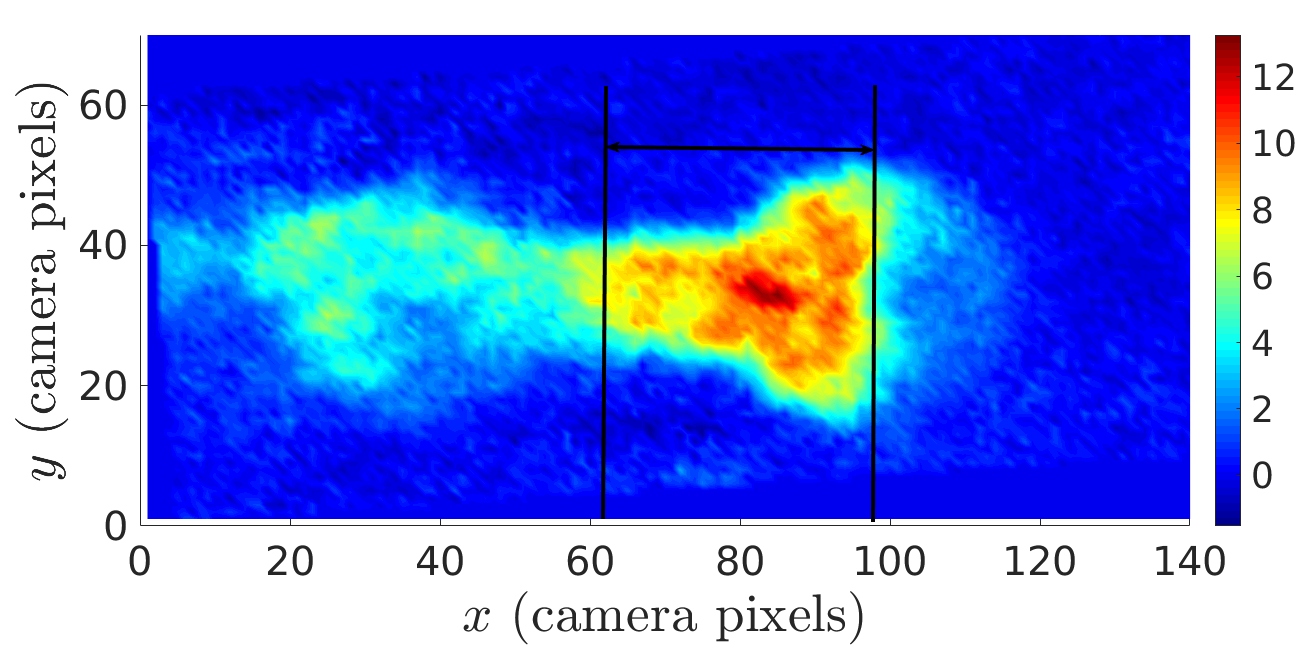}
\caption{\label{fringe} A raw experimental image (plotting the number of atoms on each camera pixel) taken at 70 ms after expansion began with no scatterers in the channel. The system size in this example is as follows: the channel width is 50 SLM pixels (36 $\mu$m), the channel length is 75 SLM pixels (54 $\mu$m), and the reservoir radius is 60 SLM pixels (43.2 $\mu$m). The BEC begins in the right-hand-side reservoir and flows to the left. The black vertical lines and the double arrow indicate roughly the extent of the dark fringe discussed in the text. For times $t>70$ ms, the images do not change significantly in terms of the outline of the fringe. This allows one to deduce the positions of the fringe edges in the source reservoir and inside the channel. The observed edges depend on the geometry of the fringe, the acceleration, and the atomic energy distribution.} 
\end{figure}

First of all, the two-dimensional (2D) trap potential is not completely uniform: it features large and deep fringes that are periodic but widely separated. The exact mechanism giving rise to these is unknown, although the authors of \cite{BS} provide a plausible theory for their origin (see section I of the Supplementary Material of \cite{BS}, where most of the discussion of the fringes is to be found, with exceptions indicated explicitly below). The fringes are mentioned in \cite{BS}, but the information given is incomplete and partially incorrect. In particular, the authors claim the fringing effect is weak, the 2D trap is nearly flat, and transport in the horizontal plane is practically ballistic (section \textit{Methods: Experiment} of \cite{BS}). These claims are false. As I will show in the rest of the Comment, the fringe depth is higher than the atomic energy so that the 2D trap does not look flat to the atoms, and motion is nowhere near ballistic, as the fringe completely traps the atoms until a significant acceleration is added (see below). Even after tilting the 2D trap, about half of the atoms remain trapped in the fringe, in contrast to an explicit statement in \cite{BS} on the matter. The position of the fringe relative to the dumbbell is also incorrectly specified in \cite{BS}: the centre of the fringe is not superimposed on the centre of the source reservoir, but on the channel opening, as can be easily seen from the raw data (see Fig.~\ref{fringe}).

Furthermore, the authors argue that the fringe has no effect on Anderson localisation physics, and that simulations performed without these potential features can be meaningfully compared to the data. I disagree with this statement. The fringe affects the density profiles, where its signature is almost always distinctly visible (Fig.~\ref{TrapConfig}). It also determines the acceleration one has to apply, the value of which appears significantly larger than quoted in \cite{BS}. It is well known that acceleration weakens Anderson localisation (see, e.g., \cite{MyBaby}) and can destroy it completely. As for the simulations presented in \cite{BS} (claimed agreement with which is used to argue that the fringe is of little importance), several parameters can be adjusted to make the outcome of the simulations more similar to the experiment (details given later), but true agreement is extremely difficult to attain -- even if the fringe is modelled -- due to the lack of knowledge about the fringe, the acceleration, the 2D trap landscape (see section \ref{topology}), the height of the scatterers, as well as uncertainty regarding the initial condition. In my opinion, the simulations have little in common with the experimental results, as will be discussed in the rest of the Comment.

I would also like to note that the published experimental results alone do not allow one to test whether data taken on different days can be compared quantitatively (minor adjustments to the apparatus were performed each day before running experiments). In other words, we cannot be sure that the fringe realisation, positioning and tilting is reproducible to a sufficiently high degree from day to day. Naturally, if such day-to-day variability is significant, it would not be possible to meaningfully model the entire collection of data, even if the fringe was included in the simulations. It would also imply that ordered and disordered experiments cannot be meaningfully compared unless the runs are performed on the same day, without adjusting the 2D trap in between.

Because the details of the fringe, the necessary acceleration, and the atomic energy distribution are extremely important for modelling and interpreting the experiment, I now provide all the available experimental information on these topics. The loading into the 2D trap from the CO$_2$ laser dipole trap is optimal if the former is positioned such that the atoms are within one of these dark fringes \footnote{Dr.~Thomas A.~Haase and Dr.~Dylan J.~Brown, personal communication (electronic mail, available for viewing upon request).}. This is precisely what was done, positioning the spatial light modulator (SLM) potential such that one of these dark fringes ran perpendicularly to the channel and spanned half of the source reservoir and the first 50 SLM pixels (36 $\mu$m) of the channel (see Fig.~\ref{fringe}).

Now, when the 2D trap is horizontal, the fringe completely traps the atoms such that without the SLM potential superimposed, the atoms simply expand to fill the fringe [5],\cite{DonaldPhD} (along the $y$ axis in Fig.~\ref{fringe}). With the repulsive SLM potential in place, the atoms are confined to the area where the dumbbell and the fringe overlap. This implies that the height of the SLM potential and the depth of the fringe are both greater than the atomic energy, because practically no atoms escape.

Acceleration is added by tilting the 2D trap, which lowers the ``channel end'' of the dark fringe sufficiently to allow about half of the atoms to escape. The ``reservoir end'' of the fringe is in turn raised, and serves to sharply stop the atoms from moving up-stream. An experimental image of the atoms in the tilted fringe (after sufficient expansion to outline its shape) is shown in Fig.~\ref{fringe}. Recall that a camera pixel has 2.1 $\mu$m-long sides at the atomic plane \cite{BS}, allowing one to calibrate figures using camera pixels as a unit into more meaningful physical dimensions if desired. Note that the edges of the fringe as seen in Fig.~\ref{fringe} depend on the geometry of the fringe, the acceleration and the atomic energy distribution. It is quite possible that a second dark fringe was encountered down-stream of the atoms for the longest channels used, but due to the acceleration, it would not trap the atoms [5]. However, I should say that there is no sign of this happening in the experiment (if it did, I would expect the SLM potential to drop so far below the atomic energy, that the dumbbell would fail to trap the atoms).

\begin{figure*}[htbp]
\subfigure[]{\includegraphics[width=3.5in]{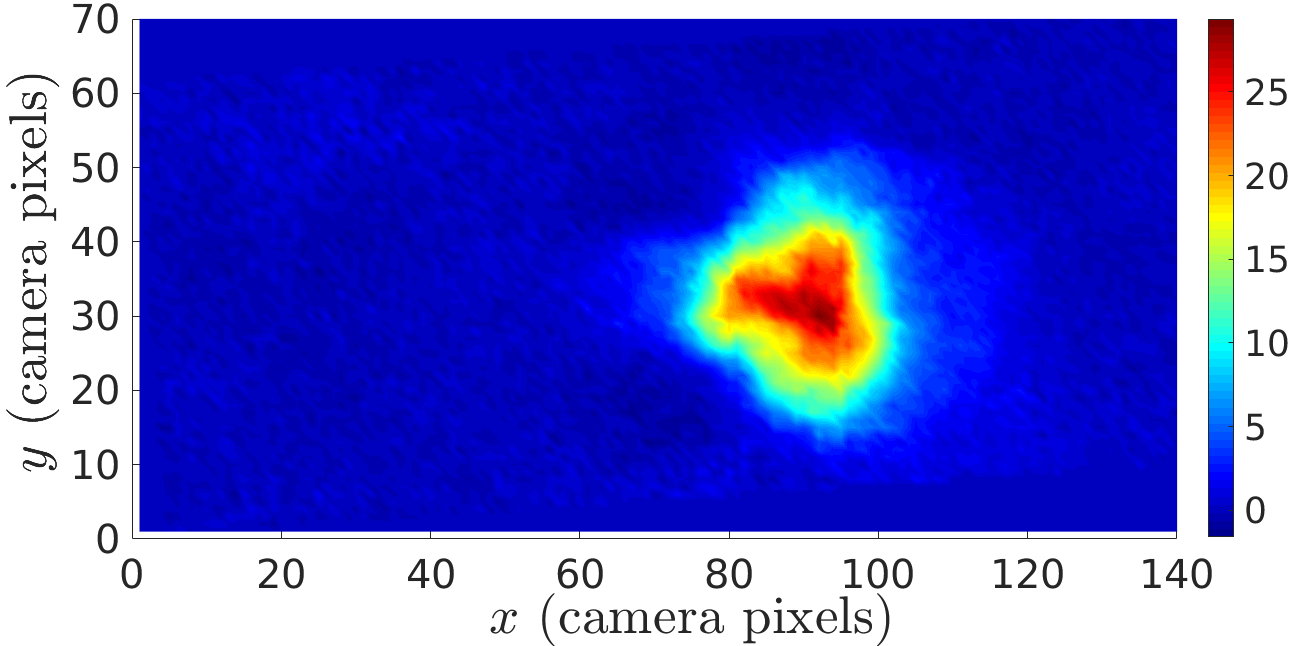}}
\subfigure[]{\includegraphics[width=3.5in]{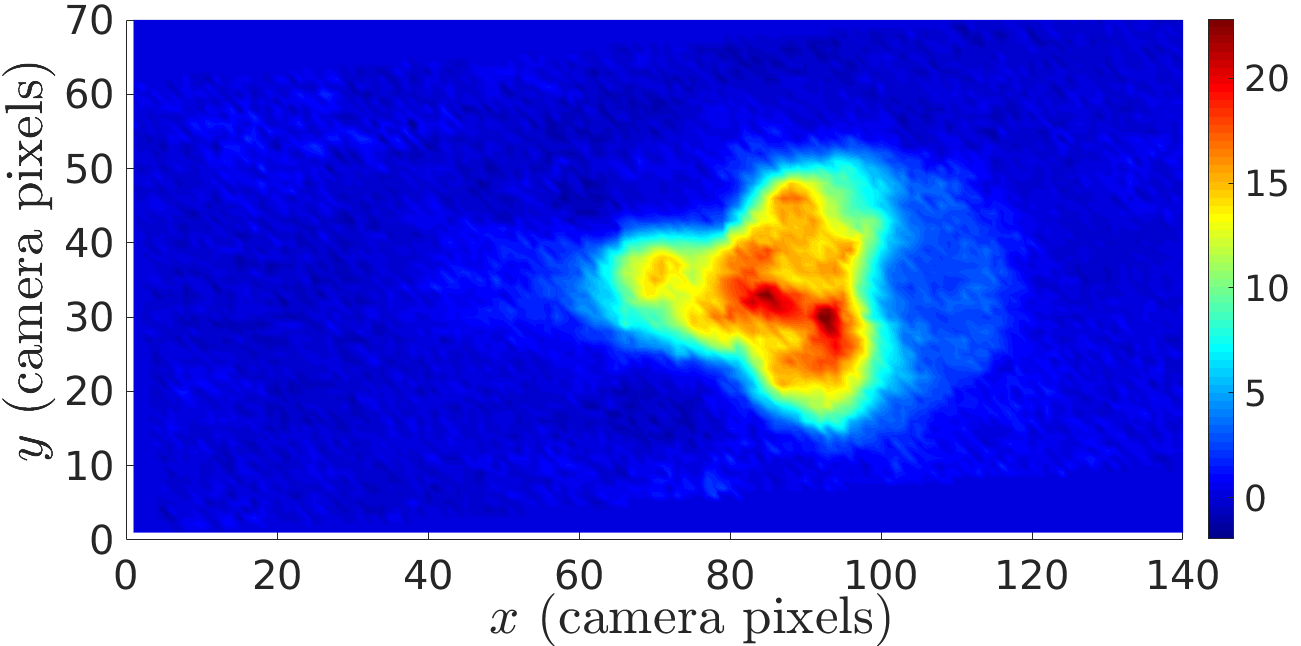}}
\caption{\label{IC} Raw experimental images (showing the atom number on each camera pixel) taken at 10 and 20 ms (panels (a) \& (b), respectively) after expansion began with no scatterers in the channel. The system size in this example is as follows: the channel width is 50 SLM pixels (36 $\mu$m), the channel length is 75 SLM pixels (54 $\mu$m), and the reservoir radius is 60 SLM pixels (43.2 $\mu$m). The BEC begins in the right-hand-side reservoir and flows to the left. These earliest-available absorption images in the time sequence can be used to test whether a proposed initial condition for theoretical modelling appropriately fits the expansion dynamics seen in the experiment. One must be aware that the potential landscape the atoms experience has a non-negligible influence, however.} 
\end{figure*}

The initial position of the Bose-Einstein condensate (BEC) was set to lie about half way between the centre of the source reservoir and the channel opening (see Fig.~\ref{IC}). Note that this implies that the minimum of the atomic energy distribution lies slightly above the minimum of the fringe as the atoms are loaded on the fringe slope. The tilt of the 2D trap was adjusted so that the fastest atoms arrived at the far end of the drain reservoir 100 ms after the CO$_2$ laser dipole trap was turned off, using a 200 SLM pixel (144 $\mu$m) long channel. Both of these facts are stated in the Supplementary Material of \cite{BS} (section I), but the main text erroneously claims the BEC is loaded at the source reservoir centre, while the numerical simulations of \cite{BS} place it at the channel opening (Supplementary Material of \cite{BS}, section IX) to compensate for the effect of the fringe which is not included. At 10 ms from the moment expansion begins, the diameter of the atomic cloud is about 60 SLM pixels (43.2 $\mu$m), and at 20 ms, it expands sufficiently to fill up the fringe. This is illustrated in Fig.~\ref{IC} and the information can be used to gain insight into the initial condition for the evolution in the 2D trap (see section \ref{Model}). However, one must realise that the potential landscape the atomic cloud evolves in impacts the observed expansion.

Note that the depth of the fringe is quoted as roughly 5 nK in \cite{BS} and the period as 150 $\mu$m (208 SLM pixels), which are reasonable order-of-magnitude estimates, but no more than that. Any serious attempt at modelling the experiment would require the full information given above. The acceleration is estimated as 0.002 m/s$^2$ in \cite{BS} (main text of \cite{BS}, section \textit{Methods: Experiment}, and Supplementary Material of \cite{BS}, sections I and IX), which appears too low to be able to match the experimental observations described in this subsection (this statement is based on attempted modelling described in section \ref{Model}). At the very least, it seems the acceleration should be higher \textit{locally} (in the vicinity of the fringe), but it is possible that other features on the 2D trap potential (e.g. the harmonic confinement or the fringe pattern itself, which may have broad, gentle maxima between the deep, dark fringes) alter the shape of the overall potential over large distances. The value 0.002 m/s$^2$ was obtained by fitting compartment population dynamics (mostly discussed in section \ref{LCR_sec}), excluding the fringe, so it can indeed underestimate the true acceleration significantly. The authors commend themselves (Supplementary Material of \cite{BS}, section IX) on their inclusion of the 2D harmonic potential (the centre of which is arbitrarily positioned in the simulations, see section \ref{HO_centre}) and the acceleration, but extremely important features such as the fringe and the details of the 2D trap landscape (see section \ref{topology}), strongly influencing the outcome of any given experiment, are left out, which brings the relevance of their simulations into question.
\subsection{Centre of harmonic trap}
\label{HO_centre}
\begin{figure*}[htbp]
\includegraphics[width=7in]{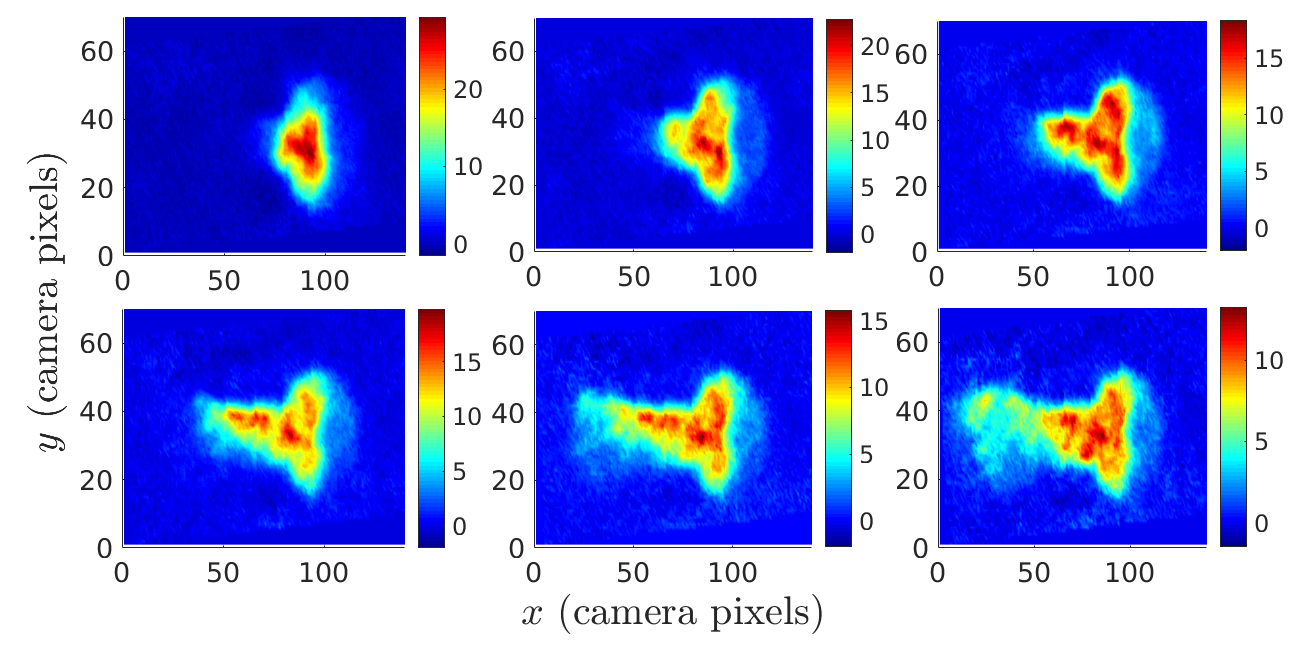}
\caption{\label{skewed} A time series of raw experimental images (plotting the number of atoms on each camera pixel) taken at 10-60 ms in steps of 10 ms (going across and down the panels) after expansion began with no scatterers in the channel. The system size in this example is as follows: the channel width is 50 SLM pixels (36 $\mu$m), the channel length is 75 SLM pixels (54 $\mu$m), and the reservoir radius is 60 SLM pixels (43.2 $\mu$m). The BEC begins in the right-hand-side reservoir and flows to the left. Note the skewed atomic flow: the preference to flow through the upper part of the channel in this fashion is observed in almost all the data sets, and is better visible in the initial surge of the atoms down the channel. In later images, such as those shown in Fig.~\ref{fringe} and later times, this preference weakens.}
\end{figure*}

Note that the experimentalists do not know with any certainty where the centre of the 2D harmonic trap lies in the plane. However, there is some indication that it is located outside of the dumbbell, on the order of 100 SLM pixels (72 $\mu$m) from the entrance to the channel, and off to one side of the longitudinal axis of the dumbbell (at least 50 SLM pixels (36 $\mu$m) away). This hypothesis is put forward based on the fact that the atoms initially flow through the empty channel in a skewed fashion practically every time, suggesting a curved potential. This skewed flow is less visible at later times. An example of such motion is shown in Fig.~\ref{skewed}. Due to lack of knowledge, in the simulations of \cite{BS} the centre of the 2D trap is placed at the centre of the dumbbell, and is maintained at the centre of the channel for different channel sizes. In the experiment, the position of the source reservoir is fixed with respect to the 2D trap position, and the channel centre moves with respect to the 2D trap centre as the channel geometry is varied. 
\subsection{Scatterer height}
\label{Scat_height}
The question we would like to address in this section is how high are the potential scatterers placed in the channel during the disordered experiments. The energy of the atoms relative to the scatterers determines how strongly the atoms are scattered and thus the degree of Anderson localisation we can expect. It also determines to what extent the scatterers are able to classically block the atoms, so this is a very important parameter, indeed. Unfortunately, it is not experimentally known \footnote{We can try to estimate the height of the background SLM potential by computing the dipole potential associated with the green beam that creates it. This sort of calculation yields about 45 nK, but this value seems overly high. It is not inconceivable that a large portion of the green beam is in fact lost, so this calculation should be regarded as an upper bound for the actual height of the SLM potential.}, so the best we can do is list some relevant information.

The potential produced by the SLM at the atomic plane is a convolution of the point spread function of the imaging system \cite{ThomasPhD,Guys2017} with the ``object'' at the face of the SLM. As such, an infinitely large object will have some ``background'' SLM potential value (depending on the intensity of the green beam), but small objects (a few pixels wide) will have their shapes significantly altered by the convolution: they will be ``washed out'', and their amplitude will be smaller than the background value. Performing the convolution both in $x$ and in $y$, a $2\times 2$ SLM pixel scatterer drops to 65\% of the background value (for this particular imaging system). Therefore, if we could deduce either one of the background SLM potential or the scatterer height, the remaining number could be easily calculated.

Incidentally, after imaging, the scatterers are similar to Gaussian bumps in shape, but in \cite{BS}, they are modelled as pillars of a constant square cross-section. While I do not expect the different shape of the scatterers to have a very strong effect, I have not attempted to quantify it thoroughly (as I simply used Gaussian scatterers in my own simulations).

Now, during the initial surge of the atoms down the empty channel, the most energetic atoms arrive at the drain reservoir first. We see a fraction of these atoms escaping across the far end of the drain reservoir both for short and long channels. The fraction of atoms lost increases with channel length (visual inspection suggests that it is so, but regrettably, I have not found a good way of quantifying this statement), but remains fairly modest. This gives us at least some link between the background SLM potential (added onto the spatially-changing 2D trap potential) and the energy distribution of the atoms that escape the fringe (this distribution has a low energy cut-off determined by the turning point of the fringe edge in the channel). We can also be certain that the background SLM potential traps the entire energy distribution even at the minimum of the fringe in which the atoms begin (see section \ref{FAIC}).

Unfortunately, this is essentially all we know. I have attempted using various combinations of these pieces of information to determine the SLM potential height, but these have led to conclusions that are incompatible with other known / assumed facts about the system, so none of these attempts can be trusted (all involve strong assumptions). The biggest difficulty is the fact that we do not truly know what the initial condition is (see section \ref{Model}), nor do we know the 2D trap potential to which the SLM potential is added. In particular, we do not know the exact acceleration, the details of the fringe pattern, or the position of the harmonic potential itself. Since the atoms -- the energy distribution of which is highly important -- are moving in the sum of the 2D trap and SLM potentials, separating the two is very challenging, especially with no certainty regarding the initial condition. Thus, we do not have an experimentally-determined value for the scatterer height.

In \cite{BS}, the scatterer height is roughly estimated as 5 nK (without providing experimental justification, but I think it is a reasonable order of magnitude), the same estimate that is given for the atomic energy and the fringe depth (see main text of \cite{BS}, section \textit{Methods: Experiment}, and Supplementary Material of \cite{BS}, sections I and IX). How one can then neglect the fringe in modelling and interpretation is unclear to me. Moreover, the background dumbbell potential created by the SLM is taken as 22 nK in the simulations of \cite{BS} (Supplementary Material of \cite{BS}, section IX), which is inconsistent with the known link between the size of the potential structure and its height at the atomic plane after imaging. The scatterer height is an extremely important parameter for numerical simulations, strongly affecting localisation properties (see, e.g., \cite{MyBaby}). In the simulations of \cite{BS}, it is a \textit{fitted} parameter. Depending on which data set is fitted, different values of the scatterer height are obtained, because many other important features that are present in the experiment are missing in the modelling. In particular, I am referring to the fringe (see section \ref{FAIC}) and the smaller-scale, but highly important structure on the 2D trap potential discussed in section \ref{topology}. Instead, the scatterer height is used to compensate for all of these features, which makes it impossible to fit all the data at once, or in a meaningful way.
\subsection{Ordered scatterer runs}
A single ``regular lattice'' experiment is shown in \cite{BS} (Supplementary Material of \cite{BS}, section VIII), comparing density profiles in a channel with disordered scatterers to one with the same number of regularly spaced scatterers. In the latter case, Anderson localisation clearly could not be at play. The preset fill-factor used was 0.16, but taking into account scatterer overlap, it was quoted in the article \cite{BS} as 0.13.

I would like to point out that the ordered run shown was performed about a month after the corresponding disordered data was taken. As mentioned previously, the published data alone is insufficient to determine whether the 2D trap setting can be accurately reproduced, so we cannot be sure that the differences visible between the ordered and disordered data are caused by the arrangement of the scatterers rather than the 2D trap potential landscape the atoms move in (including the details of the fringe).

\begin{figure*}[htbp]
\includegraphics[width=7in]{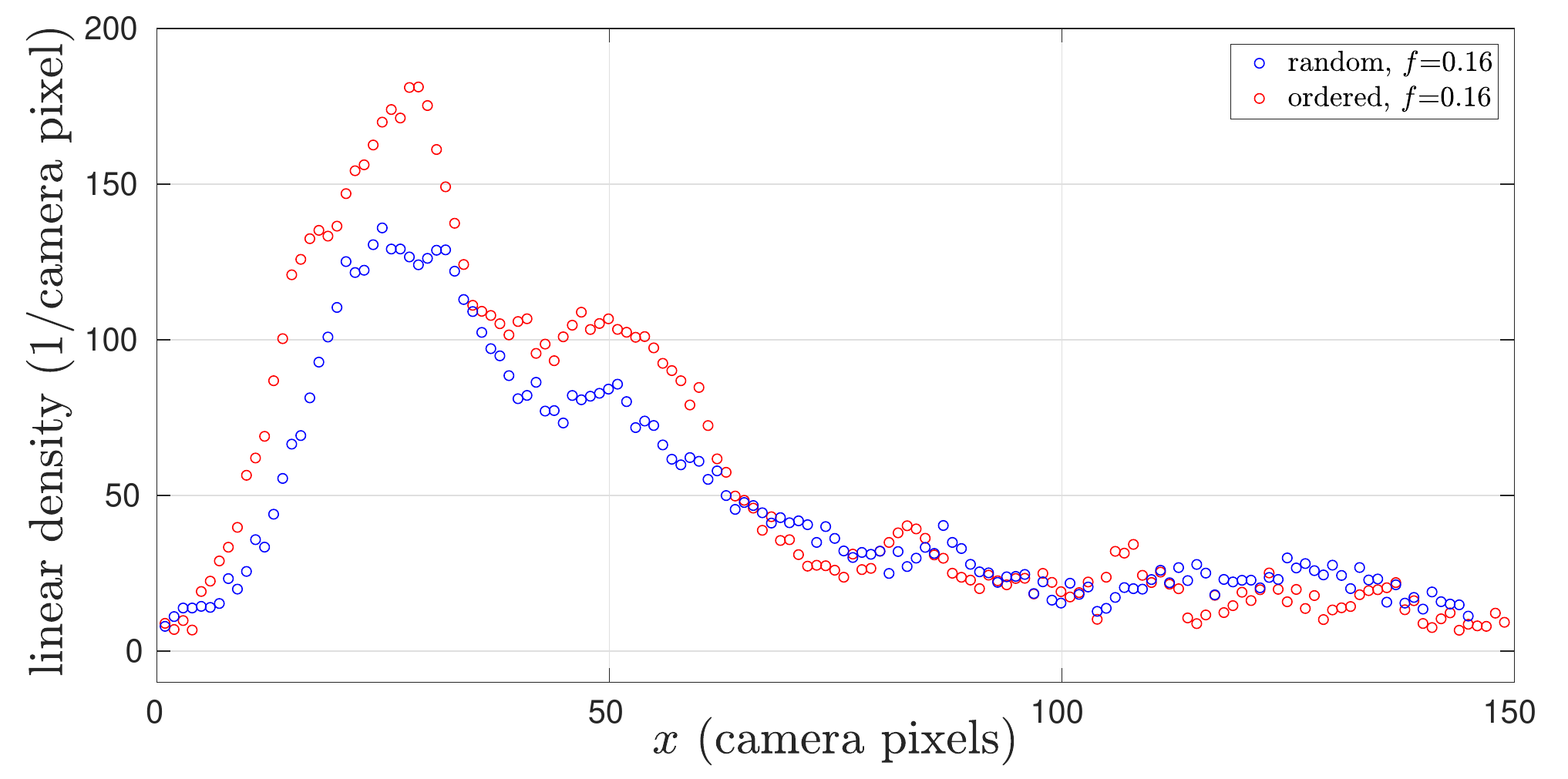}
\caption{\label{kink} Longitudinal density profiles, integrated over the width along the entire dumbbell, with the source reservoir on the left and the drain on the right, for an ordered vs.~disordered experiment. The preset fill factor is 0.16 (the value quoted in \cite{BS} is 0.13, which accounts for scatterer overlap), and the profile corresponds to 300 ms after expansion began. The channel length was 225 SLM pixels (180 $\mu$m), the width 50 SLM pixels (36 $\mu$m), and the reservoir radius 60 SLM pixels (43.2 $\mu$m) for both cases. This figure displays the same data as Fig.~11 (a) of the Supplementary Material of \cite{BS}. I fail to see a striking difference in the features, apart from that caused by the fringe, which ends where the plateau around 50 camera pixels begins to fall off. One would need to see an analogous comparison between two identical experiments performed at around the same time as these two runs, so that one could judge to what degree the 2D trap landscape can be reproduced and accurately re-aligned on different days.}
\end{figure*}

In the article \cite{BS}, it is claimed that the disordered run showed a more exponential density profile while that of the ordered run had a distinct ``kink'' in the middle of the channel and a slower-than-exponential decay at the mouth of the channel. In my opinion, the visible differences are mostly due to the particular position, shape and tilt of the fringe. The ``kink'' in the density profile is the edge of the fringe (see Fig.~\ref{kink}), which for some reason was stronger and sharper in the ordered run than in the analogous disordered one. The biggest difference between the two density profiles is seen immediately outside the fringe, after which the two density profiles merge. There is no guarantee that the 2D trap setting is reproducible to a sufficiently high degree to quantitatively compare runs performed on different days.

The authors of \cite{BS} also present simulations analogous to the experimental data, shown in Fig.~11 (b) of the Supplementary Material of \cite{BS}, and state that there is a visible difference between the ordered and disordered profiles -- that the latter is more exponential than the former. I personally do not see a significant difference. The fact that both the simulations and the experimental data indicate very small differences between ordered and disordered runs is not reassuring, it simply means there is no convincing evidence at hand of Anderson localisation in either. The authors then comment that the channel resistance, extracted from the LCR model analysis (see section \ref{LCR_sec}), is 20\% higher (in simulated data) for the disordered compared to the ordered channels. This may well be, but the LCR model analysis is intrinsically flawed (see section \ref{LCR_sec}) and due to the omission of the fringe and details of the 2D trap landscape in the simulations, the results of the simulations do not tell us anything about the experiment. Finally, the authors promise that a full study of the regular versus disordered scatterer arrangement is in preparation. I am looking forward to seeing experimental results that show a meaningful difference between ordered and disordered runs, complete with proof that the two experiments can be directly quantitatively compared.
\subsection{Fluctuations}
\begin{figure*}[htbp]
\includegraphics[width=7in]{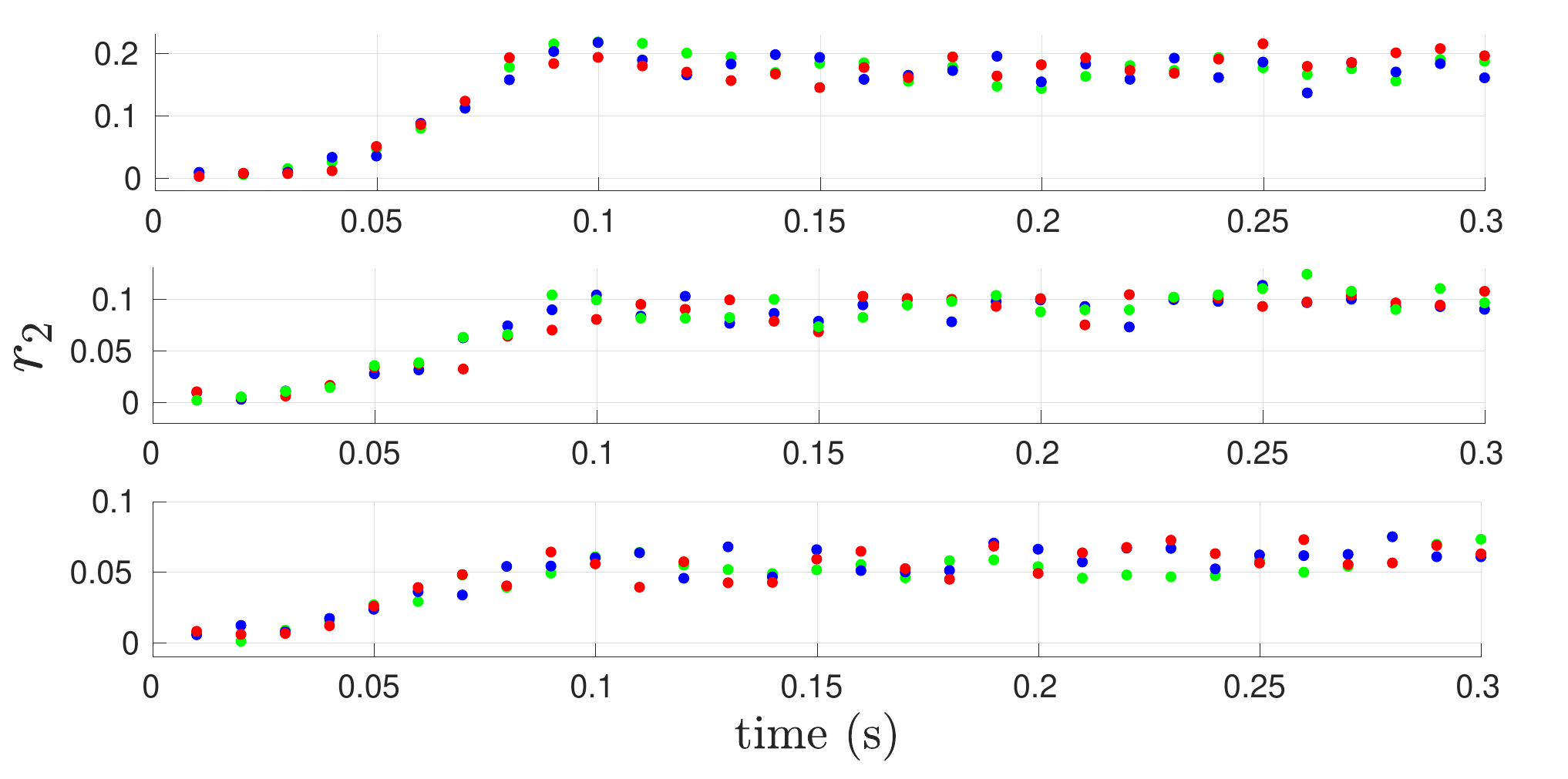}
\caption{\label{fluctuations} All panels show the population of the drain reservoir normalised by the total number of atoms as a function of time in a dumbbell with channel width of 50 SLM pixels (36 $\mu$m), length of 225 SLM pixels (162 $\mu$m), and reservoir radius of 60 SLM pixels (43.2 $\mu$m). The panels, top to bottom, correspond to preset fill factor of 0.08, 0.24 and 0.4 (not accounting for scatterer overlap; in \cite{BS} they are labelled as 0.07, 0.2, 0.32, respectively, due to scatterer overlap). The three colours correspond to three different noise realisations. The flow rate -- a quantity studied extensively in \cite{MyBaby} as a measure of transport and a tool for learning about localisation properties -- is obtainable as the gradient of the initial linear rise seen in these curves and shows very little variation. Small fluctuations between different noise realisations suggest that the physics is not governed by Anderson localisation \cite{MyBaby}.}
\end{figure*}

I would like to highlight that every experiment was repeated three times, and in the cases with disorder, a different noise realisation was used for each run. The data shows very small shot-to-shot fluctuations, as is demonstrated in Fig.~\ref{fluctuations}. What is meant by ``small fluctuations'' is simply that if we have the results of one run with a particular noise realisation, then we can predict with significant confidence that other runs (with other noise realisations) will fall roughly on top of the first set of results. Being quite familiar with the simulations performed in \cite{BS}, I can report that the fluctuations therein are likewise small. This is not the case when Anderson localisation is dominating the physics \cite{MyBaby}.
\subsection{2D trap landscape}
\label{topology}
\begin{figure*}[htbp]
\includegraphics[width=7in]{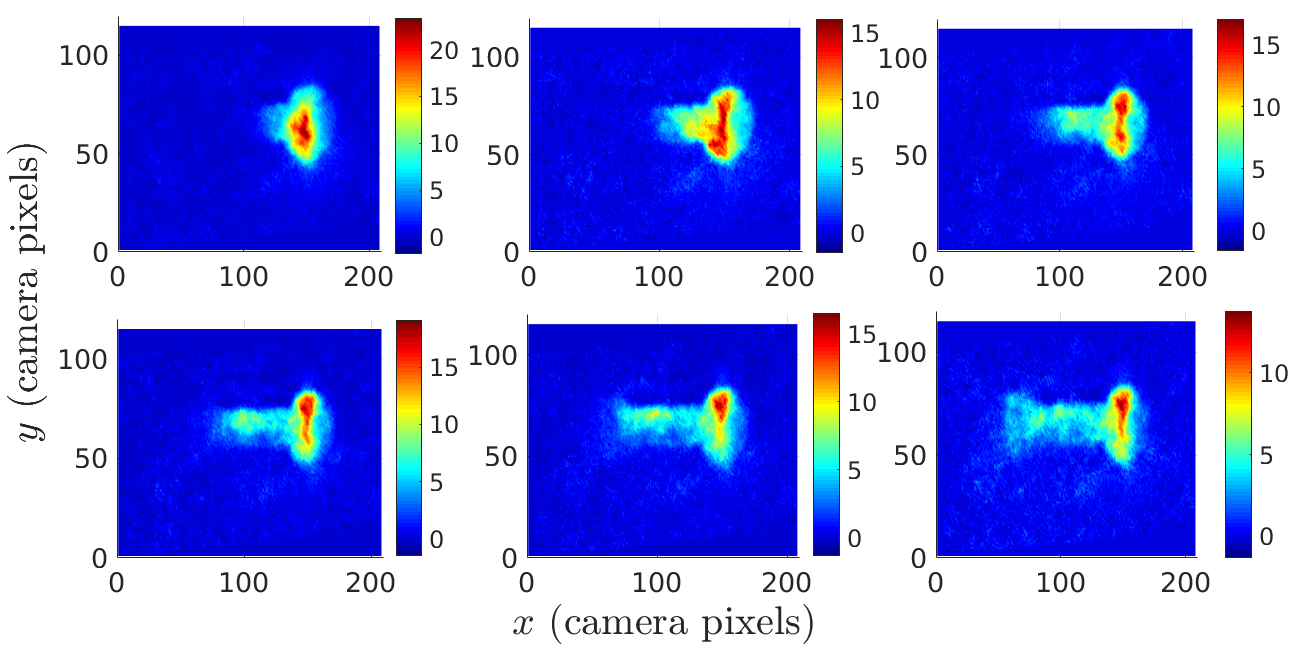}
\caption{\label{W60_2D} A time series of raw experimental images (plotting the number of atoms on each camera pixel) taken at 10-60 ms in steps of 10 ms (going across and down the panels) after expansion began with no scatterers in the channel. The system size in this example is as follows: the channel width is 60 SLM pixels (43.2 $\mu$m), the channel length is 150 SLM pixels (108 $\mu$m), and the reservoir radius is 60 SLM pixels (43.2 $\mu$m). The BEC begins in the right-hand-side reservoir and flows to the left. Note the tight confinement in the fringe inside the source reservoir. This behaviour persists for the entire duration of the experiment and of course is unaffected by the presence of potential scatterers in the channel.}
\end{figure*}

Finally, I note that the 2D trap landscape, including the dark fringes, is changeable. In particular, for most of the data collected, the 2D trap was in a \textit{qualitatively} comparable configuration, showing high reproducibility and stability over almost a month of data taking. However, some experiments (taken at the end of this month) revealed that the dark fringe can behave differently: it appears much tighter, fully confined to the source reservoir, and experiments performed in this 2D trap configuration showed completely different dynamics. For example, the associated flow rate (a quantity studied extensively in \cite{MyBaby} as a measure of transport and a tool for learning about localisation properties) was very different from what it would have been in the ``normal'' configuration, and this setting cannot be compared to the usual 2D trap setting neither quantitatively nor qualitatively. An example of the raw experimental data for such a case is presented in Fig.~\ref{W60_2D}. Furthermore, in this scenario, a drift of the fringe confinement over a several-hour time scale was observed (in this configuration \textit{only}), suggesting that the mechanism behind it is susceptible to temporal drift. The data sets taken in this 2D trap configuration are: 250 SLM pixel long and 50 pixel wide channel, 150 SLM pixel long and 40 pixel wide channel, and 150 SLM pixel long and 60 pixel wide channel, covering all investigated fill factors at these dumbbell geometries. Density profiles for three fill factors for the latter two channel geometries are shown in Figs.~2 and 3 of the Supplementary Material of \cite{BS}. A general property of this trap setting is that (apparent) localisation was weaker than in the normal 2D trap setting (density profiles in the channel look linear, at most). In fact, this is visible by eye in Figs.~2 and 3 of the Supplementary Material of \cite{BS}.

Sometimes (on other days at the end of the month over which the main data was taken), the BEC fragmented into several pieces, one circled back into the drain reservoir and one moved down the channel in a single lump (rather than spreading uniformly across the width), favouring one side of the channel every time. Note that ``lump motion'' through the channel along one side is a stronger manifestation of the skewed flow shown in Fig.~\ref{skewed}, and like the skewed flow, is observed clearly in the initial surge of the atoms down the channel. At later times, the transverse atomic density evens out somewhat and this preference to favour one side is less visible. Such lump motion has been sometimes seen in the other trap configurations as well, so it is not exclusive to this setting. The fringe appears to have been present in this trap configuration, but it was somewhat wider and shallower than in the usual trap setting, such that its effect on the density profiles was much less pronounced.

\begin{figure*}[htbp]
\includegraphics[width=7in]{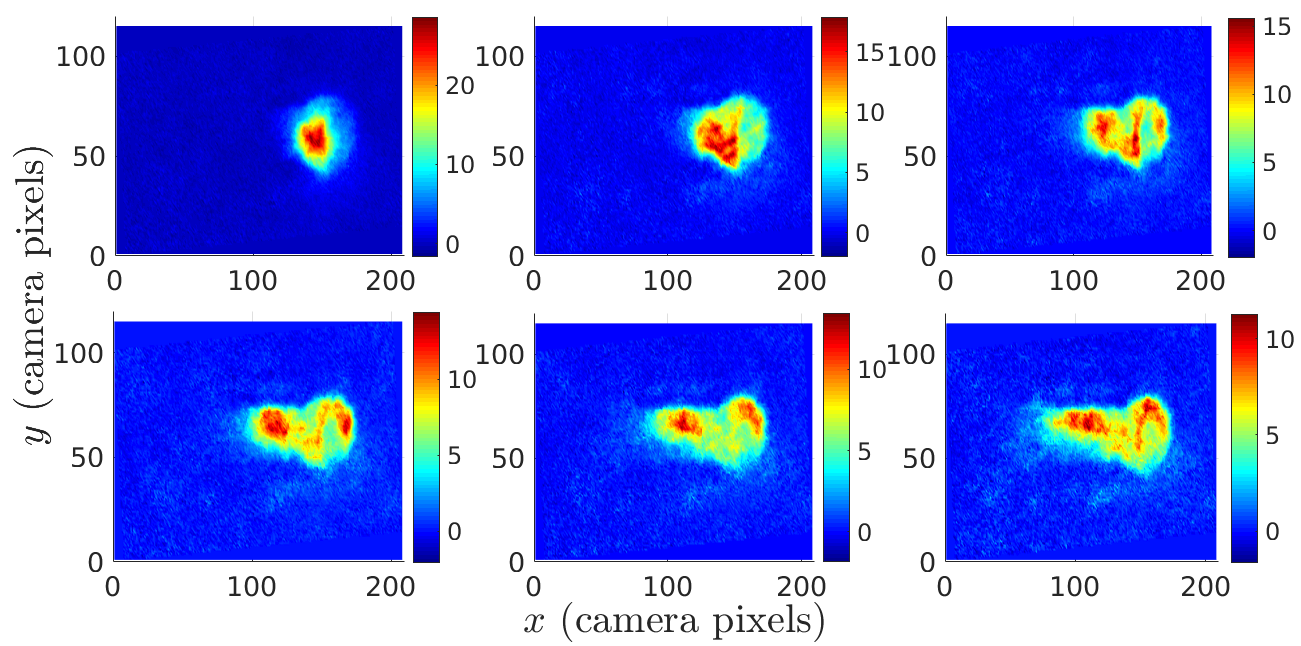}
\caption{\label{W80_2D} A time series of raw experimental images (plotting the number of atoms on each camera pixel) taken at 10-60 ms in steps of 10 ms (going across and down the panels) after expansion began with an empty channel. The system size in this example is as follows: the channel width is 80 SLM pixels (57.6 $\mu$m), the channel length is 150 SLM pixels (108 $\mu$m), and the reservoir radius is 60 SLM pixels (43.2 $\mu$m). The BEC begins in the right-hand-side reservoir and flows to the left. Note the unusual dynamics both in the source reservoir (a circling back motion is visible) and in the channel (the BEC moves as a single lump rather than spreading out over the entire width of the channel, which is even more strongly visible for wider channels in this 2D trap configuration). These are due to the rough landscape of the 2D trap at the initial position of the BEC and \textit{possibly} inside the channel (at the very least, the potential inside the channel is tilted about the longitudinal axis, and more so than in the normal 2D trap setting as the effect is more pronounced).}
\end{figure*}

There were four data sets taken in this 2D trap configuration. The channel length was fixed at 150 SLM pixels and the width was taken as $20, 80, 100, 120$ SLM pixels. Each data set (defined by a dumbbell geometry) contained several fill factors. Density profiles for three fill factors for these four channel geometries are plotted in Figs.~1, 4, 5, 6 of the Supplementary Material of \cite{BS}. An example of the behaviour of the BEC in 2D is shown in Fig.~\ref{W80_2D} for the 80 pixels width data, and it is even more pronounced for the two higher widths. In particular, it almost appears that the atoms start off at a local maximum of the potential landscape and roll down in either direction (towards the source reservoir and towards the channel). The potential landscape is certainly ``crumpled'' inside the source reservoir, as evidenced by the long-time atomic distribution therein. One can always see a region close to the centre of the source reservoir which the atoms never fill, confirming that indeed there is a potential maximum near the initial position of the BEC. A general property of this trap setting is that (apparent) localisation was much stronger than in the normal 2D trap setting (density profiles look smoother, fall off rapidly and have more curvature). This is visible by eye in Figs.~1, 4, 5, 6 of the Supplementary Material of \cite{BS}. The 80 SLM pixel (57.6 $\mu$m) wide channel data shown in Fig.~2 of \cite{BS} belongs precisely to this family of 2D trap configurations, and the claim of Anderson localisation in this article is based on exactly this data.

These observations imply that there are smaller features on the 2D trap than the fringe -- the landscape is not smooth. Sometimes, when the initial position of the BEC coincides with a ``rough'' region of the 2D potential, their effect is directly observable. Most of the time it is not, but even then one cannot rule out the possibility that such features are present in other locations in the system such that their effect is unnoticeable.

Upon presenting the various different channel width data sets in section II of the Supplementary Material of \cite{BS}, the authors point out that steady-state exponential density profiles are obtained in all cases. What they fail to comment on is the extremely strong effect the particular fringe and 2D trap landscape realisation (which are completely out of the experimentalists' control and may well change from day to day) have on the dynamics and the density profiles. In fact, the authors go on to say that the 20 SLM pixel wide channel appears to be blocked by what they describe as ``minor'' imperfections and inhomogeneities in the 2D trap landscape. The channel is certainly blocked in this case, but the features on the 2D trap responsible for this are far from ``minor'' and their effect dominates the outcome of experiments. To a certain degree, they attribute the poor transport into the narrowest channel at zero fill factor to reflections off the edges of the channel at its opening. However, I have performed many simulations, which show that the dynamics observed in the experiment have nothing to do with reflection at the mouth of the channel. The only way (that I could find) to explain the experimental data for the 20 pixel wide channel is the proposition that the channel is blocked by features on the 2D trap landscape. Thus we know that there are rough features on the 2D trap landscape near the mouth of the channel for the wider channel data-sets as well (80, 100, 120 SLM pixels), because these were all taken in the same 2D trap configuration. However, with much wider channels, the atoms can simply go around these features, and so the effect is much less noticeable. 
\section{Numerical modelling of the experiment}
\label{Model}
In this section, I briefly outline the procedure one would have to follow to numerically reproduce the experiment \cite{BS} with brute-force Gross-Pitaevskii (GP) simulations. Specific challenges involved in such an endeavour are numerous. For one, the shape, depth and width of the fringes on the 2D trap are unknown -- one can only try to determine these by matching the observed effect of the fringe on the atoms. Two simple shapes that can be used are a sinusoidal modulation and a box-like ``trench'' with variably-smoothed edges. Second, the exact value of the acceleration is unknown. It can be estimated by comparison of simulations to experiment, but a large number of other factors need to be set correctly at the same time as they influence the BEC's motion just as much. These are the atomic energy distribution, the fringe shape, depth and width, and the position of the 2D trap centre. Incidentally, in simulations performed by both the authors of \cite{BS} and myself using initial conditions with realistic energy, the atoms always moved down the channel noticeably slower than in the experiment, which suggests a stronger acceleration in the latter. Third, the height of the repulsive SLM potential is also not known, and can only be gauged relative to the atomic energy from the observed effects. Needless to say, the rough features of the 2D trap seen in the trap configurations discussed in section \ref{topology} cannot be modelled as we know nothing about them.

Unfortunately, the initial condition is also not known exactly. Approximately, one could hope that the initial wavefunction is a 2D Thomas-Fermi (TF) profile arising from the CO$_2$ laser dipole trap that originally holds the BEC, before it is released into the 2D trap by abruptly removing the CO$_2$ laser beam. Consider the experimental procedure: after creation of the BEC through evaporative cooling in the three-dimensional (3D) CO$_2$ laser dipole trap, the latter is slowly ramped down while the 2D trap potential is slowly ramped up (which squashes the atoms into a pancake-shaped cloud). If the ramps are sufficiently slow for the process to be adiabatic, the end result is a wavefunction in the ground state of the combined CO$_2$ laser dipole and 2D trap potentials. Since the $z$ frequency of the 2D trap is much higher than the $z$ frequency of the CO$_2$ laser dipole trap, and vice versa for the transverse confinement, we can (approximately) neglect the weaker harmonic trap in each direction. Under these assumptions, when the CO$_2$ laser dipole trap is abruptly switched off to release the atoms, we would indeed obtain a TF wavefunction. The simulations performed in \cite{BS} created a 3D BEC in the ramped-down CO$_2$ laser dipole trap and integrated it across the $z$-dimension to reduce it to 2D for time-evolution. However, the presence of the 2D trap was not included in obtaining the initial condition. Ideally, full modelling of the 2D trap loading procedure is desirable to correctly compute the true initial condition used in the experiment (or at least check the validity of the TF approximation). Note that if we use the TF profile, we should ensure the wavefunction only has support in a region over which the 2D trap potential is flat -- otherwise, potential features such as the fringe must be included already when solving for the initial state.

Let us assume for the moment that the initial condition is indeed a TF profile. Can we match the size and expansion rate based on the first two available images (see Fig.~\ref{IC})? Note that the precise CO$_2$ dipole trap frequency at the end of the ramp is unknown and is simply guessed for the simulations of \cite{BS}, with no attempted fitting. We can use the analytical solution for a 2D TF cloud expanding into an empty, infinite 2D plane \cite{Kamchatnov} (which I have tested against exact numerical simulations \cite{MyBaby}). We can fix the particle number at $N=15,000$, as in all the experimental data sets, the initial total number of atoms is consistently between 14,000 and 16,000. Using an initial TF radius of 18 SLM pixels (13 $\mu$m) leads to radii of 33.2 SLM pixels (24 $\mu$m) and 58.4 SLM pixels (42 $\mu$m) at 10 and 20 ms, respectively. This is not incompatible with the observations. On the other hand, an initial TF radius of 25 SLM pixels (18 $\mu$m) leads to radii of 32 SLM pixels (23 $\mu$m) and 47.2 SLM pixels (34 $\mu$m) at 10 and 20 ms, respectively. This choice can also not be ruled out from the two initial snap-shots, as the radius at 10 ms is almost the same and by 20 ms, the dumbbell and fringe walls come into play and constrict the expansion. I remark that such attempts at matching the experimental images naively assume that the potential landscape the atoms are moving in can be neglected.

Expansion data is available from an earlier version of the experiment \cite{Guys2017}, but the radius there is extracted from a Gaussian fit to the cloud profile, and a green cylinder is used \cite{DonaldPhD} to overcome the fringing, so there is no guarantee that these numbers can be taken on face value. Furthermore, significant improvements were made to the apparatus between the point when the data in \cite{Guys2017} and in \cite{BS} was taken. In short, there is ambiguity regarding the initial condition as well, and the difference between the two initial TF radii mentioned above -- 18 and 25 SLM pixels -- changes the energy distribution significantly. Incidentally, in the asymptotic limit of a TF wavefunction expanding in 2D, the energy distribution falls off linearly and the maximal energy in the distribution is 3 times the energy of the initial state \cite{Kamchatnov}. Not knowing the initial condition adds yet another layer of complexity to fitting the experiment.

To summarise, many experimental features and parameters are unknown, and attempting to fit all of them at the same time is a monumental task. This problem is not easily circumvented as all of them strongly affect the experimental observations.
\section{Criticism of Ref.~\cite{BS}}
\subsection{Experimental arguments}
\label{Crit_Expt}
While I have no hard proof because I have never managed to properly reproduce the experiment in a numerical simulation, intuition developed in the course of the investigation presented in \cite{MyBaby} suggests that it is unlikely that Anderson localisation was observed in \cite{BS}. The reasons for that are:
\begin{itemize}
\item If the scatterers are indeed of the order of 5 nK as estimated in \cite{BS} (which does not seem unreasonable to me), they are quite low compared to the atomic energy. Low scatterers imply weak localisation, as I have shown repeatedly in \cite{MyBaby}.
\item The atoms appear to be rather energetic, in the sense that my attempts at modelling the experiment with the lower-energy TF profile mentioned above were unsuccessful and it appeared to me that a more energetic BEC would be a step in the right direction. Energetic atoms are more difficult to localise because the localisation length increases with energy.
\item Acceleration seems relatively strong (at least in the vicinity of the fringe), which can be judged from how much the fringe needs to be tilted to let half of the atoms flow out. Of course this depends on the specifics of the fringe width, shape and depth, and the atomic energy distribution, but a rough estimate can be made nonetheless. As I have shown in \cite{MyBaby}, acceleration has a detrimental effect on Anderson localisation.
\item The comparison of ordered scatterers to disordered experiments would only be significant if the authors can prove that the 2D trap can be re-aligned from day to day with sufficient precision for the \textit{same} experiment to be reproducible to a degree where the differences between \textit{different} experiments can be cleanly ascribed to the nature of the experiments.
\item In addition, each experiment was performed using three different noise realisations, and the observed fluctuations were small. The significance of this piece of evidence is that in the course of my theoretical investigation \cite{MyBaby}, I have found that small shot-to-shot fluctuations in the flow rate (see \cite{MyBaby}) indicated that the dynamics are not governed by Anderson localisation.
\item A further complicating factor is that the fringe extends over the first 50 SLM pixels (36 $\mu$m) of the channel, such that over this distance, the entire energy distribution is able to probe the disorder, while downstream of the fringe, only the higher end which is able to escape from it (certainly not advantageous to observing localisation).
\end{itemize}
\subsection{Interactions}
\label{Ints_sec}
Another factor that needs to be discussed in relation to whether Anderson localisation would have been observable in \cite{BS} or not are interparticle interactions. This argument is not based on experimental observations or an attempt at matching them with brute-force simulations, but rather on ``idealised'' numerical simulations (neglecting the fringe, for example), and is thus presented separately.

Repulsive interactions are at play throughout (see \cite{MyBaby}, where I have demonstrated that for a relatively low-energy TF cloud, 20\% of the energy stays locked in interactions). Interactions are detrimental to Anderson localisation, as shown in \cite{MyBaby}. In \cite{BS}, the authors state that interactions are very weak, supporting their claims with numerical simulations (Supplementary Material of \cite{BS}, section X), concluding interactions need to be five time stronger to get an observable effect. In contrast, my simulations in \cite{MyBaby} show a visible difference with the true strength of interactions already. The reasons behind the different modelling results are several: \cite{BS} places the BEC at the mouth of the channel while \cite{MyBaby} initiates the condensate at the centre of the source reservoir, the initial wavefunction is different, and the analysis performed in \cite{BS} is limited to examining the slope of a fitted straight line to the logarithm of the density in the channel.

Incidentally, it is also not specified in \cite{BS} at what time interactions are switched off (this is important as one should monitor the conversion of interaction to kinetic energy, as was done in \cite{MyBaby}). In \cite{MyBaby}, it was found that while (with the true interaction strength) switching off interactions prevents a larger fraction of the atoms from traversing the channel, the apparent, overall localisation length is unchanged. Since \cite{BS} only examined this slope, it is likely they have missed the other observations made in \cite{MyBaby} in regard to the effect of interactions. Note that estimating the average atomic density by dividing the total particle number by the total area and computing interaction energy from that (as was done in \cite{BS} for an order-of-magnitude argument) can be deceptive, because in reality, the atoms are not equally spread out over the dumbbell when the atoms traverse the channel.
\subsection{Selected data}
\label{chosen_data}
The main data displayed in Fig.~2 of \cite{BS} (showing ``strong localisation''), on the basis of which the entire claim of the paper was made, was taken with the 2D trap in the abnormal configuration shown in Fig.~\ref{W80_2D}, with a structured landscape inside the dumbbell clearly revealed by the empty channel behaviour. Amongst the published data sets (which are currently available), convincing exponential decay in the density profiles was only observed (a) in the structured trap configuration mentioned above, and (b) \textit{perhaps} at the largest fill factor (0.4 preset, or 0.32 with scatterer overlap accounted for) in long channels in the normal 2D trap setting. In the more usual, smooth configuration of the 2D trap, at fill factors below half filling, the results were much weaker than the data chosen for the main argument of the article. In my opinion, the 2D trap landscape plays a major role in determining the degree of apparent localisation seen in the disordered data. The observations in the disordered runs are a combination of the effect of the fringe and 2D trap setting, and the influence of the scatterers.

\begin{figure*}[htbp]
{\includegraphics[width=7in]{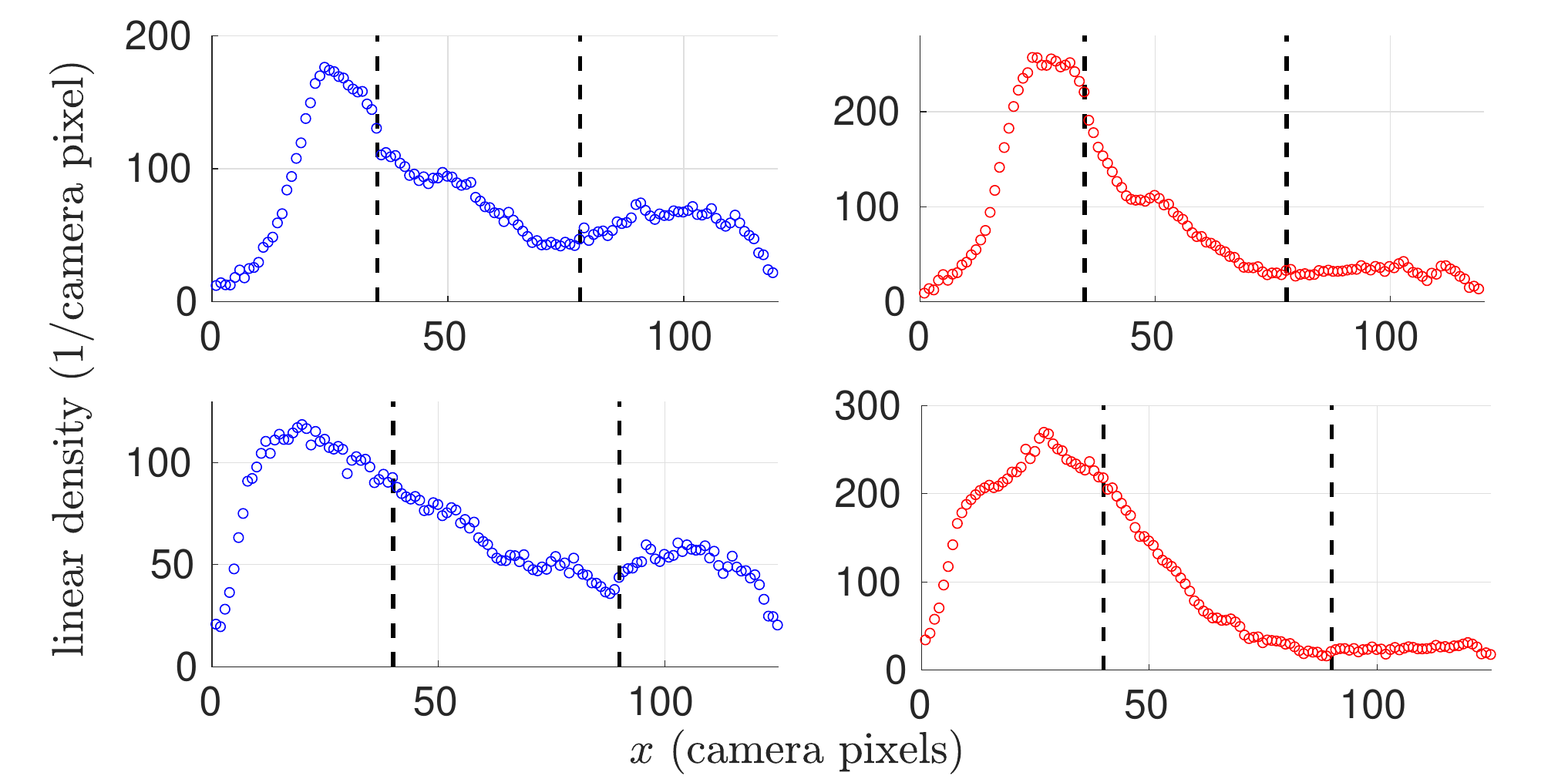}}
\caption{\label{TrapConfig} Integrated 1D density profiles at 250 ms after expansion began. Left: empty channel (bottom) and a preset fill-factor of 0.08 (0.07 with overlap accounted for) using randomly placed scatterers (top). Right: disordered scatterers, preset fill-factor of 0.4 (bottom) and 0.32 (top) [respectively 0.32 (bottom) and 0.26 (top) with overlap accounted for]. The channel length is 150 SLM pixels (108 $\mu$m), and the reservoir radius is 60 SLM pixels (43.2 $\mu$m) in all cases. Top row: the channel width is 50 SLM pixels (36 $\mu$m), bottom row: 80 SLM pixels (57.6 $\mu$m). Vertical lines show the edges of the channel. The source reservoir is on the left. In the ``normal'' trap configuration (top), the effect of the fringe is stronger and its edge leaves a clear imprint on the density at $x\approx$50 camera pixels. In the ``structured'' 2D trap setting (bottom), the fringe is (probably) shallower, so its effect is less obvious, but its channel edge can still be located (with the help of the 2D density data) at $x\approx60$ camera pixels. Overall, localisation appears stronger for the wider channel, as the entire profile can be easily taken for a single decaying curve. However, the observed ``decay'' is partially due to the 2D trap potential in this particular setting. Please note that unfortunately, since only the published data is available for presentation, in the top panels (normal 2D trap configuration), I am forced to show data with fill factor as close as possible to the analogous experiments in the bottom panels. Should \textit{all} the existing data become available, one could use the same fill factor exactly, which would change the figures quantitatively, but not qualitatively.}
\end{figure*}

Let us confirm this by examining the data directly, comparing empty channels to those filled with scatterers for the ``normal'' and ``structured'' trap settings, as shown in Fig.~\ref{TrapConfig}. The fringe on the 2D trap potential is stronger in the normal configuration, and clearly divides the profile into a section in- and out-side of the fringe. Because of the strongly changing background potential inside the fringe (which extends 50 SLM pixels (36 $\mu$m) into the channel), one cannot expect the entire density profile in the channel to be a single exponential curve in the presence of scatterers. Thus, only the latter segment should be searched for signs of localisation: a linear fall off in the empty channel develops some curvature with a high fill factor. In the structured configuration, the fringe appears to be shallower and wider, such that after tilting, its signature is almost lost in the density profile (it can still, however, be located on the 2D density plots quite readily). The linear fall-off in the empty channel density develops some curvature in the farther half of the channel (with a sufficiently high fill factor), but because the fringe is visually ``lost'', one can fit an exponential to the entire profile, a measurement which points to considerably stronger localisation than one would get in the normal 2D trap setting. This is also what Fig.~2 (e) of \cite{BS} reports.

The question now becomes whether the ``structured'' trap configuration measurements should be trusted over the ``normal'' one. That is, is it likely that the structured 2D trap configuration allowed Anderson localisation physics to dominate better than the normal configuration? If we had reason to believe that the trap setting which yielded strong localisation was smoother and more uniform, approaching idealised conditions, perhaps that would be justified. Unfortunately, as I have shown, there is actually evidence pointing to the contrary. The 2D potential landscape was quite uneven, with smaller rough features than the fringe inside the source reservoir, near the channel mouth, and (possibly) in the channel, though the fringe itself was probably more ``gentle''. Taking this into account, as well as the fact that one may never be able to reproduce this particular trap configuration (which heavily contributes to the observed density profiles) ever again, I would think that taking these results on face value is ill-advised. Furthermore, fluctuations in the transmission in the ``structured'' setting were not noticeably higher than in the ``normal'', implying that the observed stronger localisation was not due to disorder. 

None of the information and discussion in this subsection is provided in \cite{BS} at all, and the difference between panels (a) and (b), (c) of Fig.~2 is entirely attributed to the fill factor and a ``mysterious'' width dependence (see section \ref{width_sec}). I remark again that in the Supplementary Material of \cite{BS}, density profiles at different channel widths are shown, with four cases being in one abnormal 2D trap configuration, and two in another (see section \ref{topology}). The difference in the density decay caused by the landscape of the 2D trap and the fringe behaviour is immediately obvious by eye. The article \cite{BS} contains an extensive discussion attempting to rule out classical trapping as the cause of the observed density decay. However, these arguments assume a flat 2D potential with scatterers of known density and height, whereas in reality, there are features on the 2D trap landscape that are as high or higher than the scatterers, and on different spatial length-scales: the fringe is a large-scale structure, while the features of the structured landscape on the 2D trap are smaller-scale. Thus, the arguments against classical trapping are not relevant for the experimental data shown.
\subsection{Width dependence}
\label{width_sec}
Two statements are strongly implied in \cite{BS}: first, that the strong exponential decay in the experiment is always seen at fill factors above 0.2 (preset), and second, that the apparent significant decrease seen in the localisation length with increasing width extracted from the experiment is real. The first has been thoroughly discussed in the previous subsection (at a fill factor of 0.2 in the normal trap configuration, there is no question of strong localisation or convincingly exponential density profiles at all, e.g.~see Fig.~\ref{kink}). The second statement is, in my opinion, not true. The authors justify the second claim by displaying GP simulation results which show that the extracted localisation length is smaller for a wider channel. On the other hand, the drop in localisation length in simulations is much smaller than in experiment, where it is very strong but the authors describe it as ``slight''.

Furthermore, the GP simulations in \cite{BS} were fitted to the wide channel data (used to claim localisation) and without adjusting the scatterer height, show no agreement with the narrower channel data presented (see Fig.~2 (a) \& (e) of \cite{BS}). The error bars on the experimental data in Fig.~2 (e) of \cite{BS} for the longer channel are so large because first of all, the standard deviation is plotted rather than the standard error, and second, the exponential fits are very poor because the density is not truly exponential (see sections \ref{chosen_data} and \ref{Exp_fits}). As for Fig.~2 (a) of \cite{BS}, the authors note the disagreement between theory and experiment, but reassure the reader that if one waits much longer in simulations, the density profiles do eventually become nearly flat (shown in Fig.~8 of the Supplementary Material of \cite{BS}). The reason that it takes longer in simulations is because the scatterer height is larger in simulations (among many other discrepancies) than in experiment, as it was fitted to the wider channel data and compensates also for the strong effect of the uneven 2D trap landscape (the narrower channel data was taken in the more ``normal'' 2D trap configuration). The authors acknowledge that there are factors missing from the simulations, including ``minor'' deviations from flatness in the 2D trap, but they \textit{significantly} understate their importance and effect.

The authors of \cite{BS} do not understand the width dependence seen in their simulations (nor do they provide an explanation for the experimental observations), hypothesising that these are due to finite size effects. This is not the case. The difference in the localisation properties seen in the experiment is completely due to the 2D trap configuration which utterly overshadows and dominates over any finite size effects one might have seen had the 2D trap been flat. In fact, the true theoretical dependence of Anderson localisation on the channel width has been extensively studied in \cite{MyBaby}, revealing a monotonic growth of the localisation length with increasing width, in agreement with many years of previous work by other authors (see \cite{MyBaby} for a review). The theoretical prediction (coming out of the simulations of \cite{BS}) of a channel width at which the localisation length abruptly drops is most-certainly not due to finite size effects, but may be an artefact arising from how the data was processed, at best.

The authors of \cite {BS} proceed to promise a full future study of the width dependence seen in the experiment and in their simulations. I am \textit{eagerly} awaiting to see this forthcoming publication, especially since there was ample time (and motivation) to investigate these matters prior to the publication of \cite{BS} itself.

I have personally thoroughly and methodically studied the width dependence via numerical simulations. These efforts include the results presented in \cite{MyBaby}, which point to nothing but a trivial increase in the amount of matter flowing through the channel as the width is increased, accompanied by a \textit{weakening} of intrinsic localisation properties due to finite-size effects. My findings are consistent with several decades of previous literature on the subject (see \cite{MyBaby}). In addition, I have spent several months running extensive GP simulations, including all the secondary features mentioned in \cite{MyBaby} (going above and beyond what is included in the simulations presented in \cite{BS}), and confirmed via direct, brute-force modelling that there is no mechanism in existence that would cause Anderson localisation to strengthen upon increasing the width -- it simply does not happen. I have confidence in my results due to the testing of the codes performed (see appendices of \cite{MyBaby}) and the fact that \textit{several} theoretical approaches all give the same prediction. I do not know how to explain the drop in localisation length seen in the GP simulations presented in \cite{BS}, but suspect it could be an artefact of how the numerical data is processed. The authors should certainly investigate what is happening in their simulations and resolve this issue.

As for the experimental data, the 2D trap configurations (see section \ref{topology}) are fully responsible for the otherwise-inexplicable changes in transport properties. These have nothing to do with Anderson localisation as the same strong differences are seen for empty channel runs. This last statement can be substantiated by examining the flow rate, $\rho$, which is a perfectly-well defined quantity (measuring transport properties) even in the absence of an exponential density profile. This so-called ``flow rate'' is defined and studied in \cite{MyBaby}: it is the maximal rate of the initial growth of the population of the drain reservoir normalised by the total number of atoms in the dumbbell. It is proportional to the transmission coefficient of the potential in the channel. It reports on the influx of atoms into the drain reservoir and is closely related to the longitudinal density at the end of the channel.

\begin{figure*}[htbp]
\subfigure[]{\includegraphics[width=3.5in]{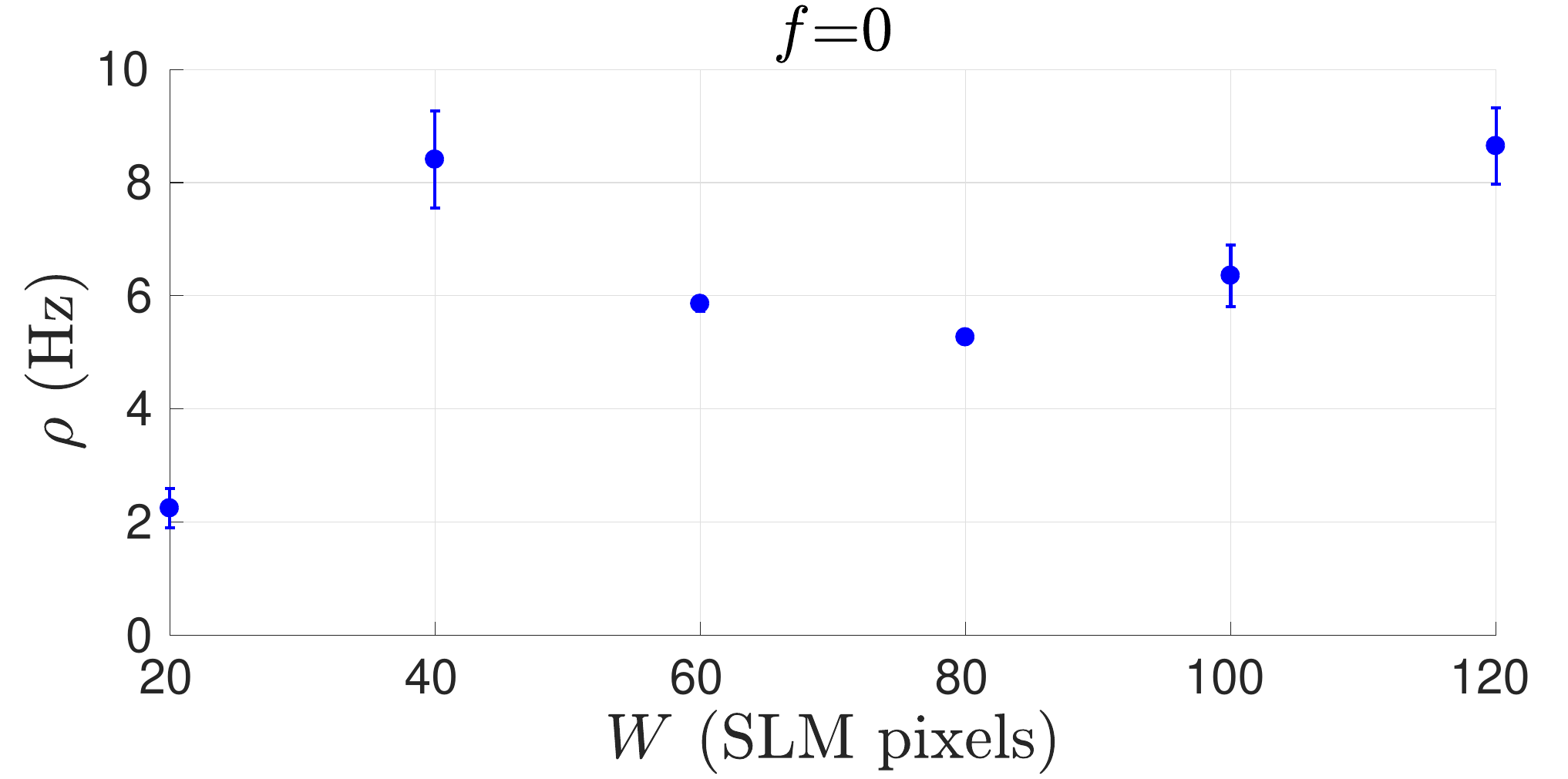}}
\subfigure[]{\includegraphics[width=3.5in]{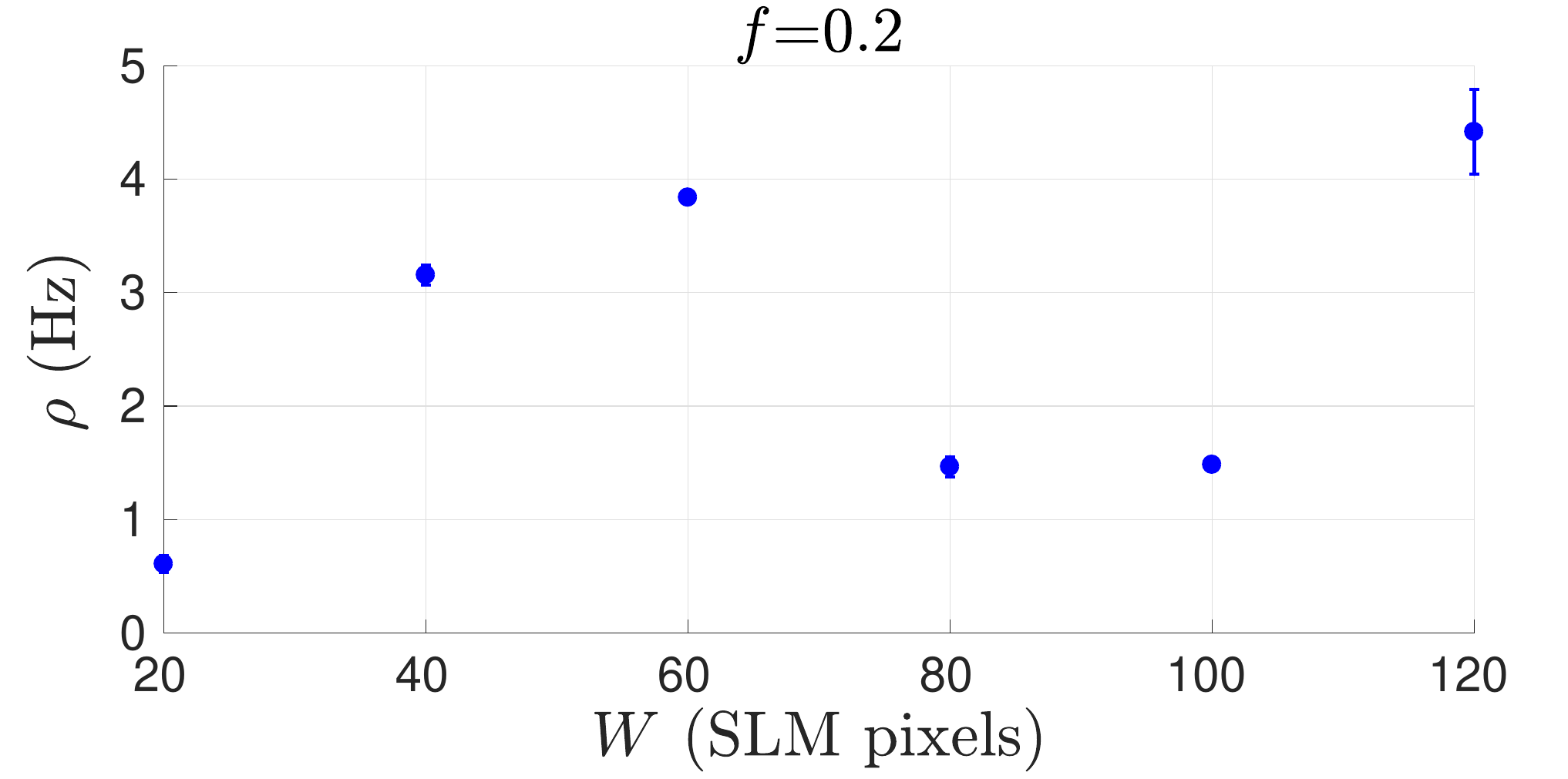}}
\subfigure[]{\includegraphics[width=3.5in]{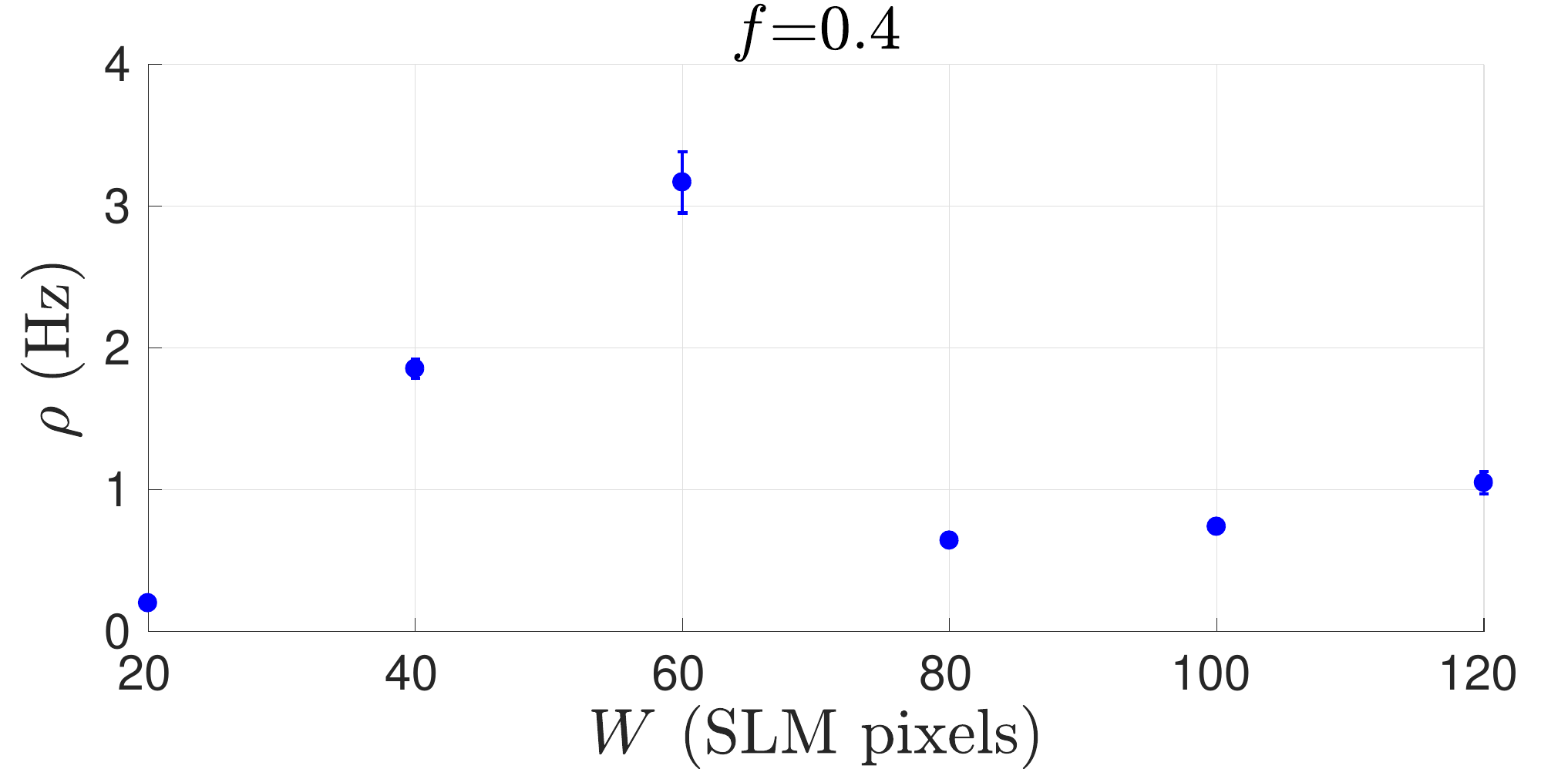}}
\caption{\label{widthdep} The flow rate out of the channel as a function of channel width for different (preset) fill factors $f$ indicated above each panel. (Accounting for scatterer overlap, the fill factor for the second and third panels was quoted in \cite{BS} as 0.17 and 0.32, respectively.) The channel length was 150 SLM pixels (108 $\mu$m), and the reservoir radius was 60 SLM pixels (43.2 $\mu$m) for all experiments. Error bars depict the standard error. The overall trend seen at fixed fill factor is that the flow rate increases from 20 to 60 SLM pixel widths, drops abruptly, and then continues to increase slowly. This strange pattern is explained by the different 2D trap configurations that were realised when the 40 and 60 SLM pixel width data was taken, and that which was realised for the other four data sets. Within a 2D trap configuration family, by and large, increasing the width increases the flow rate, as it should. Exceptions (outliers) to these trends can be seen in the data: for the empty channel, transport appears to have been better at $W=40$ compared to $W=60$ SLM pixels. This is explained by the drift of the fringe confinement on the day these two width data sets were taken. Looking at the raw data, one can see the fringe traps the atoms much more strongly and sharply for 60 pixels, hence the decrease in the measured flow rate. This tighter confinement was also an issue when scatterers were present in the channel, but here the overall trend is an increase in $\rho$ despite it. As for the $W=120$ SLM pixels data point with preset fill factor $f=0.2$, it seems to lie overly high relative to the other points. However, we remark that the $W=80,100$ SLM pixel data sets were taken on one day while the $W=20,120$ SLM pixel data sets on another, which means that despite the fact that all four are in a qualitatively comparable 2D trap configuration, daily variability in the specific realisation and the tilt of the fringe could potentially cause large fluctuations. To summarise, the 2D trap configuration has a strong effect on the measured transport properties and explains the ``mysterious'' width dependence.}
\end{figure*}

The relevant plots are displayed in Fig.~\ref{widthdep}. The flow rate for the ``structured'' 2D trap configuration at which the $20, 80, 100, 120$ SLM pixel wide channel data was taken is much smaller than that of the data taken in the ``tight-fringe'' 2D trap configuration (channel widths $40, 60$ SLM pixels) at \textbf{all} fill factors, including zero. Within the same 2D trap configuration, the flow rate always increases with growing channel width, barring the case when the fringe confinement significantly drifted during the day (one can easily see a strong difference between the behaviour of the fringe during the 40 and 60 SLM pixel wide channel data sets). The normal 2D trap configuration shows intermediate transport (and apparent localisation) properties, lying between these two abnormal trap configurations.

In Fig.~7 of the Supplementary Material of \cite{BS}, the authors show the localisation length extracted from the experimental data for the three widest channels used (80, 100, 120 SLM pixel width) at two fill factors (0.2 and 0.4 preset, or equivalently, 0.17 and 0.32 with scatterer overlap accounted for). All this data was taken in the same 2D trap configuration, where apparent localisation was much stronger than in the other trap configurations. Indeed, one does not see a significant dependence of the extracted localisation length on the channel width. However, if the results for the other three channel widths (20, 40, 60 SLM pixels) are added to this figure, then a huge variation is visible: qualitatively, the plot resembles the bottom panel of Fig.~\ref{widthdep}, with the relative change on the y-axis very much of a similar order of magnitude (I do not present this figure as I did not do the relevant calculation myself, but merely saw the final result). The authors claim that the smallest three widths are omitted from Fig.~7 of the Supplementary Material of \cite{BS} because in that range, ``finite size effects are strong''. The strong variation seen in the experiment comes from the 2D trap configuration being different for the data sets with different widths \footnote{In addition, recall that the 20 SLM pixel wide channel was classically blocked by features on the 2D trap.}. Finite size effects \textit{may} have been visible if the experimental system was ``clean'' (a truly flat 2D trap, weak or no acceleration, no interactions, etc.) and the scatterers were high compared to the atomic energy (see \cite{MyBaby} for a relevant discussion), but as things stand, the 2D trap landscape is completely dominating the ``effective localisation length''. This is clear also because of the fact that finite size effects predict a monotonically increasing localisation length with system width, rather than the exotic shape seen in the bottom panel of Fig.~\ref{widthdep}.
\subsection{The LCR model}
\label{LCR_sec}
A less crucial (yet still important) criticism of \cite{BS} is the use of the LCR model to extract a ``channel resistance'' for the disordered dumbbells. Previously, three groups have used this model, and each has implemented a somewhat different version of it: \cite{Esslinger2012, Esslinger2012b} used a simple LCR circuit with no added complications, \cite{Eckel2016} found they needed to add a weak-link to the model in order to fit the data, and \cite{Edwards2016} opted for a strongly time-dependent resistance instead. In my opinion, this model is poorly physically motivated, with the analogy to an actual LCR circuit being quite weak. Nevertheless, in principle, it is possible that this model would actually capture the behaviour of the atomtronic system correctly. Unfortunately, this does not appear to be the case, at least for the results of the experiment at hand (see below).

\begin{figure*}[htbp]
\includegraphics[width=7in]{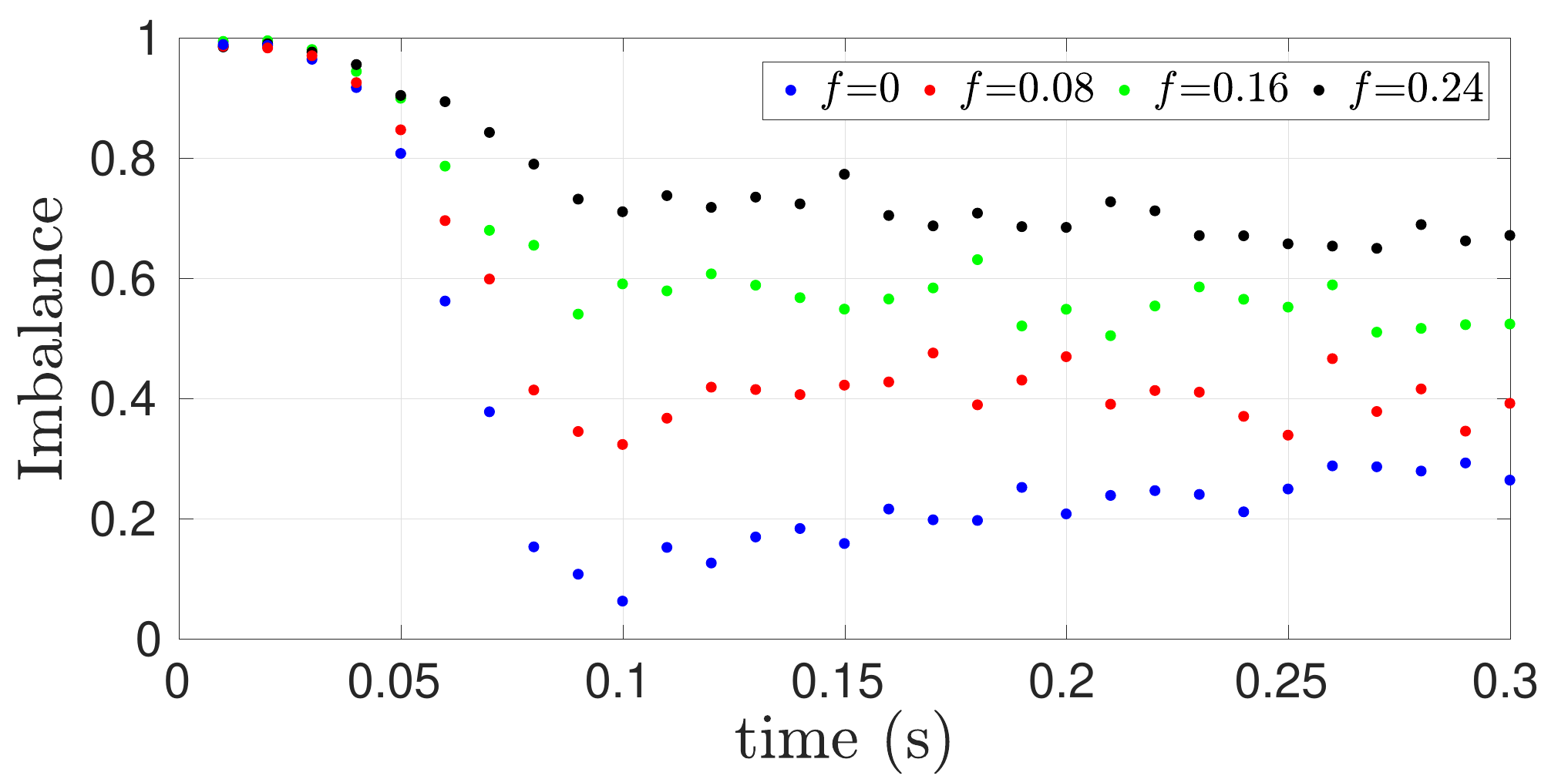}
\caption{\label{LCR} The imbalance as a function of time for a dumbbell with channel width of 50 SLM pixels (36 $\mu$m), channel length of 225 SLM pixels (162 $\mu$m), and reservoir radius of 60 SLM pixels (43.2 $\mu$m). The preset fill factor is indicated in the legend. The same data is shown on a semilogarithmic plot in Fig.~4 of \cite{BS} (where the fill factor is quoted accounting for scatterer overlap). The data points between 50 to 100 ms certainly do not appear to be exponentially decaying.} 
\end{figure*}

The simple LCR model used in \cite{BS} implies that the imbalance (see \cite{BS} for the definition) will decay exponentially with time. There is only a very short period of time in the experimental data where this behaviour could even \textit{qualitatively} be suspected of being applicable, and even then it is not convincing at all when the data is examined on a linear scale. This is demonstrated in Fig.~\ref{LCR} for the exact same data plotted in Fig.~4 of \cite{BS}. Next, the authors of \cite{BS} comment that the channel resistance is related to the transmission coefficient of the channel. This, in my opinion, is unjustified. There is no clear connection that I can see. Meanwhile, the flow rate observable defined in \cite{MyBaby} \textit{is} directly proportional to the transmission coefficient.

Finally, I remark that the imbalance in the experimental does not usually vanish at long times with zero fill factor (see Fig.~\ref{LCR}). This is due to the significant fraction of the atoms (about half) that remain trapped in the fringe, roughly half of which contribute to the population of the source reservoir and half to that of the channel, simply due to the position of the fringe relative to the dumbbell in these experimental runs (see section \ref{FAIC}). Indeed, the presence of the fringe has a strong influence on the compartment population dynamics, and a mathematical model which excludes it cannot truly reproduce the experiment. This is one reason why in the simulations of \cite{BS}, the BEC was initiated at the channel opening (see Supplementary Material of \cite{BS}, section IX) rather than at the true initial position -- to compensate for the effect of the fringe on the compartment population dynamics.
\subsection{Exponential fits}
\label{Exp_fits}
\begin{figure*}[htbp]
\subfigure[]{\includegraphics[width=3.5in]{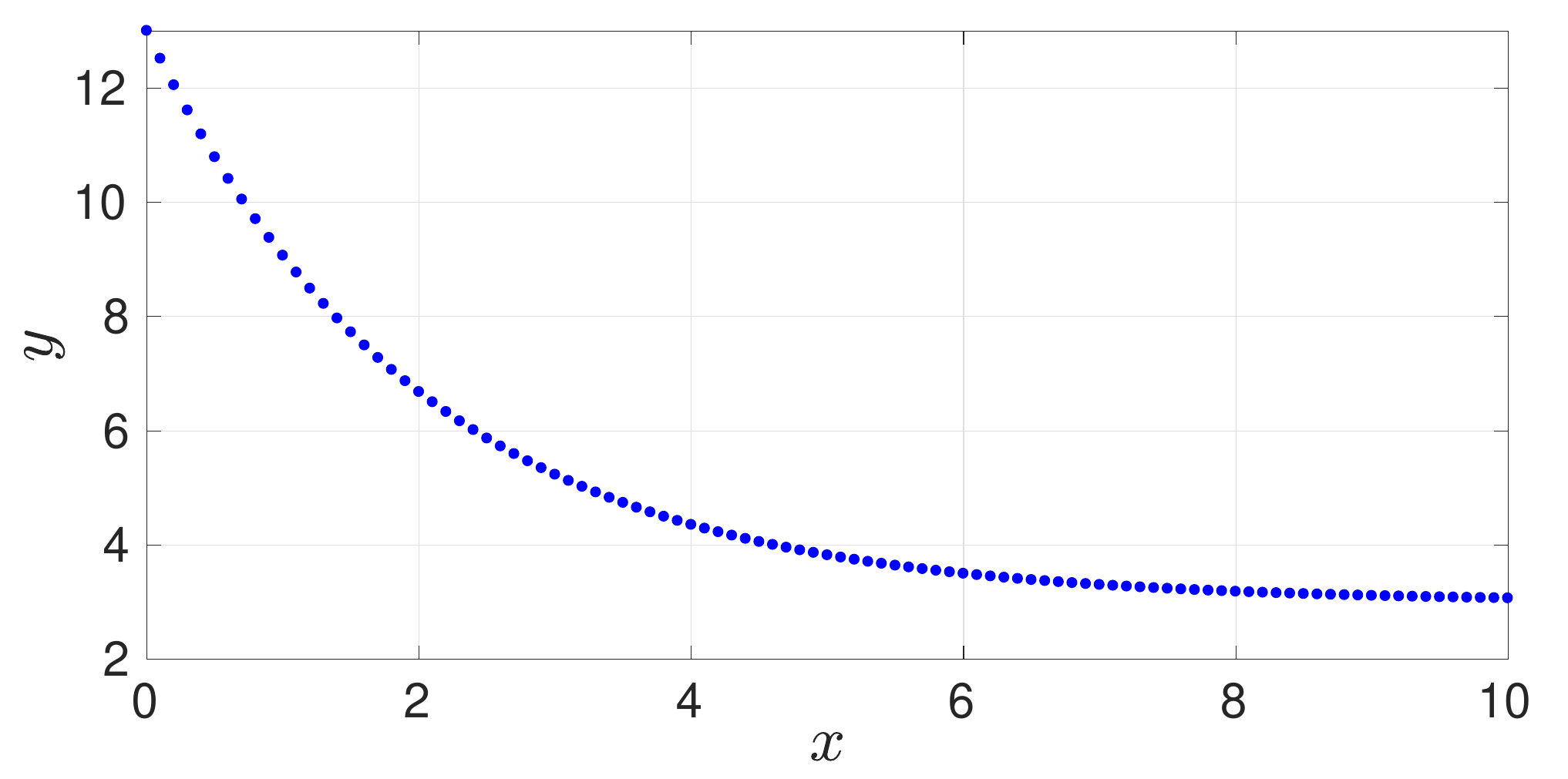}}
\subfigure[]{\includegraphics[width=3.5in]{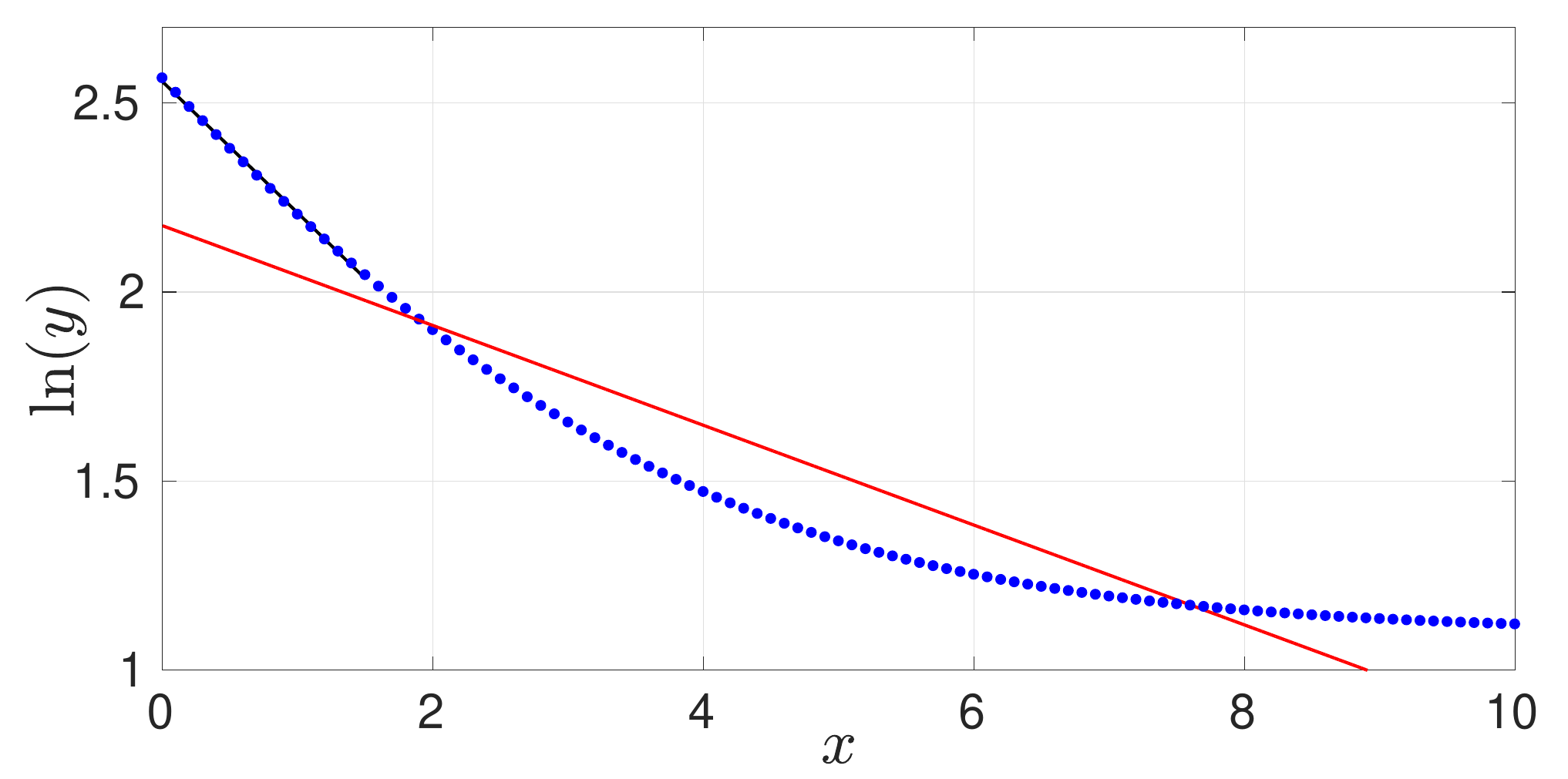}}
\caption{\label{maths} The function $y = 10\exp(-x/2)+3$ is plotted (in blue) directly (left) and after taking the natural logarithm (right). While the dependence of $y$ on $x$ is without question exponential, due to the off-shift from zero, $\ln(y)$ is not linear in $x$. Fitting the initial linear segment (black line) on the logarithmic plot yields a gradient of $-0.347$ instead of $-0.5$. Fitting all the data (red line) on the logarithmic plot yields a gradient of $-0.132$, moving even further from the true value. This highlights the problem with the way exponential fits were performed in \cite{BS} (who often followed one or the other of the two procedures).} 
\end{figure*}

All exponential fits to experimental and simulation data in \cite{BS} were performed by plotting the logarithm of the data and fitting a straight line to it. Mostly, this was done rather carelessly, without differentiating segments of the density profiles that look exponential from those that do not \footnote{This statement can be tested by looking into the details of the analysis of the experimental data and simulation outputs performed by the authors.}. For example, the ``step'' in the 1D density profiles in Figs.~\ref{TrapConfig} and \ref{kink} which arises from the fringe was simply ignored and the entire 1D profile over the channel was fitted as a whole. Exponential fits were performed even in cases when the density (or imbalance) did not look exponential on a linear scale, did not look linear on a logarithmic scale, and when it did not decay strongly enough within the available data range to render the off-shift negligible (this was true in most cases). If a significant off-shift is present to an exponentially decaying function, it does not look linear on a semi-logarithmic plot and the true decay rate cannot be extracted by linear fitting. A demonstration is given in Fig.~\ref{maths}. A gradient extracted this way is thus quite meaningless.
\section{Conclusion}
To summarise, I identified several important aspects that were completely neglected both in the interpretation of the experimental results and in the GP simulations presented in \cite{BS}. In the simulations, the initial condition may well be the wrong wavefunction with the wrong energy distribution (no fitting of the initial condition was attempted), placed in the wrong initial position (compare the initial position of the BEC in the experiment and in the simulations reported in \cite{BS}). The large-scale and deep fringes on the 2D trap are ignored -- both in the simulations and in the interpretation of the experiment -- as well as the smaller scale inhomogeneities about which very little is known. The centre of the 2D trap in the GP simulations was positioned without attempting to match the experiment. The acceleration was set by matching compartment population dynamics without the fringe and possibly using the wrong initial condition, which implies that the acceleration is probably incorrect. Furthermore, the scatterer shape in the simulations is incorrect while the height was matched (specifically to the wide channel data) having fixed all the other parameters and ignoring the 2D trap landscape, using the scatterer height to compensate for the missing features.

The main experimental results showing ``strong localisation'' were taken in an abnormal 2D trap configuration, with a structured landscape, which contributed greatly to the specific density profiles obtained. Other 2D trap configurations did not produce convincingly exponential density profiles at reasonable fill factors (below 0.4 preset). The ordered scatterer experiment performed for comparison was taken on a different day, and before the differences seen can be considered as significant, one would need to prove that the 2D trap was set up in a reproducible manner. Adding to this the small shot-to-shot variation between noise realisations, I doubt that the experiment actually observed Anderson localisation as is claimed in \cite{BS}. Moreover, it is very likely that the GP simulations included in the paper have almost nothing in common with what actually happened in the laboratory. All the theoretical arguments (as opposed to simulations) presented in \cite{BS} also assumed a flat background potential, which is so far from the truth of the experiment, that these arguments are simply not relevant. I have also pointed out that the data analysis performed in \cite{BS} is somewhat questionable. Finally, the width dependence is attributed to ``real physics'' in \cite{BS}, whereas in the experiment it is caused by artefacts of the 2D trap potential, and in the simulations, the origin is unknown and requires investigation.
%
\bibliography{MyRefs}
\end{document}